\documentclass[10pt,a4paper,nofootinbib]{article}
\pdfoutput=1
\usepackage{jheppub}
\usepackage[utf8]{inputenc}
\usepackage{graphicx,floatflt,amssymb,rotate, cancel}
\usepackage{tablefootnote}
\usepackage{mathrsfs}
\usepackage{amssymb}
\usepackage{amsmath}
\usepackage{bm}
\usepackage{color}
\usepackage{xcolor}
\usepackage{graphicx}
\usepackage{multirow}
\usepackage{array}
\usepackage{float}
\usepackage[toc,page]{appendix}
\usepackage{diagbox}
\usepackage[mathscr]{euscript}
\usepackage{cleveref}
\usepackage{soul}
\usepackage{pifont}
\newcommand{\cmark}{\ding{51}}
\newcommand{\xmark}{\ding{56}}
\usepackage{subfig}

\makeatletter
\gdef\@fpheader{}
\makeatother

\definecolor{orange}{rgb}{1,0.5,0}

\newcommand{\met}{\ensuremath{/ \hspace{-.7em} E_T}}

\title{Searching for exotic Higgs bosons from top quark decays at the HL-LHC}
\author[a,b)]{Gautam Bhattacharyya,}
\author[d)]{Indrani Chakraborty,}
\author[e)]{Dilip Kumar Ghosh,}
\author[e)]{Tapoja Jha,}
\author[a,b,c)]{Gourab Saha}
\affiliation[a)]{Saha Institute of Nuclear Physics, 1/AF Bidhan Nagar,
  Kolkata 700064, India}
\affiliation[b)]{Homi Bhabha National Institute, Training School
  Complex, Anushaktinagar, Mumbai 400094, India}
  \affiliation[c)]{Universite' de Strasbourg, CNRS, IPHC UMR 7178, Strasbourg, France}
\affiliation[d)]{Department of Physics and Material Science and Engineering, Jaypee Institute of Information Technology, A-10, Sector-62, Noida 201307,
Uttar Pradesh, India}
\affiliation[e)]{School of Physical Sciences, Indian Association for
  the Cultivation of Science, 2A $\&$ 2B, Raja S.C. Mullick Road,
  Kolkata 700032, India}
\emailAdd{gautam.bhattacharyya@saha.ac.in}
\emailAdd{indrani300888@gmail.com}
\emailAdd{tpdkg@iacs.res.in}
\emailAdd{tapoja.phy@gmail.com}
\emailAdd{gourab.saha@iphc.cnrs.fr}

\abstract{ Exotic spin-$0$ states with unusual couplings with the
  gauge and matter fields of the Standard Model are worth exploring at
  the CERN LHC. Though our approach is largely model independent, we
  take inspiration from flavor models based on some discrete
  symmetries which predict a set of a scalar and a pseudoscalar having
  purely off-diagonal Yukawa interactions with quarks and leptons.  In
  a previous paper, some of us explored how to decipher such exotic
  scalar and pseudoscalar states whose off-diagonal Yukawa couplings
  involve light quarks. In this work we follow a complementary path
  and focus on the Yukawa couplings that necessarily involve a top
  quark. If one such spin-$0$ state is lighter than the top quark,
  then the rare decay of the latter, on account of the high yield of
  the $t\bar t$ events, could provide a potential hunting ground of
  those exotic states particularly during the high luminosity phase of
  the LHC run. We carry out an exhaustive collider analysis of some
  promising signatures of those exotic states using sophisticated
  Machine Learning techniques and obtain considerable signal
  significance.}

\begin{document}
\maketitle

\renewcommand{\thesection}{\Roman{section}}  
\setcounter{footnote}{0}  
\renewcommand{\thefootnote}{\arabic{footnote}}  

\section{Introduction} \label{intro}

Nonobservation of any new particle so far at the CERN LHC, beyond
those which constitute the Standard Model (SM), fuels speculation that
our search strategies could be biased towards assuming standard
conventional interactions for the nonstandard states. In a previous
study \cite{Bhattacharyya:2022ciw}, some of us explored how to
decipher at the LHC the possible existence of light exotic neutral
spin-$0$ states having unconventional gauge and Yukawa interactions.
More specifically, we assumed that a relatively light pseudoscalar
($\chi$) and a scalar ($H$) exist with the following nonstandard
properties:
\begin{itemize}

\item There are no $H VV$-type couplings, where $V \equiv W^\pm,
  Z$. The $H \chi Z$ coupling takes the simple form\,($q_{\mu} \equiv$
  momentum transfer, $\theta_W\,\equiv$\,weak angle): 
  \begin{equation}
   H \chi Z~:~
  \left(\frac{-ie}{2\sin\theta_W\cos\theta_W}\right) q_\mu \,.
 \end{equation}
\item $H (\chi)$ has {\em only} flavor off-diagonal Yukawa
  couplings, with the Yukawa Lagrangian given by 
  \begin{equation}
  Y_{ff'}
  \bar{f}\,(i\gamma^5)\,f' H\,(\chi) ~+ ~ {\rm h.c.} \,.
  \end{equation}
  where, $f,f' \equiv e\mu, \mu\tau, e\tau, uc, tc, ut, ds, db,
  sb$.
  
\end{itemize} 

Although our approach would be sufficiently model independent, as also
mentioned in \cite{Bhattacharyya:2022ciw}, $\chi$ and $H$ with the
above properties do emerge as byproducts in a wide class of flavor
models which contain three Higgs doublets, e.g. those relying on
flavor groups $S_3$ \cite{Bhattacharyya:2012ze,Bhattacharyya:2010hp,Khater:2021wcx,Kuncinas:2020wrn,Kuncinas:2022whn}
or $\Delta (27)$ \cite{Bhattacharyya:2012pi}. Since there is no $HVV$
coupling and the Yukawa couplings of $H$ and $\chi$ remain purely
off-diagonal, the LEP2 limit \cite{Teixeira-Dias:2008omh} does not 
apply on the mass of $H$ or
$\chi$. Both $H$ and $\chi$ can be taken to be light. The parent
flavor models from which the above couplings descend generally contain
three Higgs doublets, i.e. they contain an additional set of neutral
and charged nonstandard scalars than offered by a two Higgs doublet
model. $H$ and $\chi$ happen to be the lighter of the two sets of
nonstandard CP-even and CP-odd scalars. Although their innate flavor
symmetries do not permit $VVH$ coupling, the $ZZ\chi$, $W^\pm H^\mp\chi$ and
$W^\pm H^\mp H$ couplings do exist. It is known in a two Higgs doublet
context, as demonstrated e.g. in \cite{Bhattacharyya:2013rya}, that if the mass
splittings between the nonstandard charged and neutral scalar states
are within 100 GeV range, the constraint from the electroweak oblique
parameters can be comfortably satisfied, provided the absolute masses
of those states weigh within a few hundred GeV.  It may be noted that
a charged Higgs weighing as low as a few hundred GeV is compatible
with the LHC data for low $\tan\beta$ with appropriate nonstandard
Yukawa couplings.  In the context of models with three Higgs doublets,
the second set of charged and neutral scalars may be kept a bit
heavier (e.g. $\sim$ TeV) and as degenerate as possible to be
consistent with electroweak precision constraints~\cite{Chakraborti:2021bpy}. 
It is also worth
noting that the two sets of charged Higgs in a three Higgs doublet
scenario may conspire to partially cancel each other's contribution to
$b \to s \gamma$. As a result, a relatively light charged Higgs
weighing a few hundred GeV can be accommodated within the radiative
$B$ decay constraints \cite{Boto:2021qgu}.

The strategies for uncovering these exotic states were chalked out in
Ref.~\cite{Bhattacharyya:2022ciw} when the relevant off-diagonal
Yukawa couplings of $\chi$ and $H$ involved only the light quarks. It
should be noted that such couplings trigger tree level meson
mixing. To avoid stringent constraints from $K^0$--$\bar{K}^0$ mixing,
the $\chi ds$ and $H ds$ couplings were set to zero. An approximate
phenomenological relationship between the ratio of $\chi uc$ and $H
uc$ couplings, as proportional to the $\chi$ and $H$ masses, was taken
to remain consistent with the constraints from $D^0$--$\bar{D}^0$
mixing. On the leptonic side, the couplings involving the electrons
were set to negligible values to avoid constraints from $e^+e^- \to
\mu^+\mu^- (\tau^+ \tau^-)$, $\mu$--$e$ conversion and $\mu \to e
\gamma$ processes. The only relevant leptonic Yukawa interactions
involved were $\chi \mu\tau$ and $H \mu \tau$. It was assumed that
$\chi$ weighs a few tens of a GeV and $H$ is at least 100 GeV heavier
than that.  The collider study involved production of $H$ dominantly
from the parton level $uc$ fusion in $pp$ collision, followed by
splitting of $H$ into $\chi$ and $Z$. Eventually, on-shell decays
$\chi \to \mu\tau$ and $Z \to \ell^+\ell^-$ constituted the final
states in Ref. \cite{Bhattacharyya:2022ciw}\footnote{For lepton flavor
  violating exotic Higgs decays at the LHC/HL-LHC, see also
  \cite{Arganda:2019gnv,Barman:2022iwj}}.

In the present paper, we perform a complementary study by focusing on
the Yukawa couplings involving the top quark, the relevant couplings
being $\Phi tj$, where $\Phi \equiv H,\chi$ and $j\equiv u,c$. We
assume any other quark Yukawa couplings to be extremely small, or
vanishing, for the following reasons. Let us recall that the
assumption of order 100 GeV neutral scalars with off-diagonal Yukawa
couplings calls for special scrutiny from the perspective of
constraints on flavor physics, in particular from meson -- anti-meson
mixing as well as from meson decays, as mentioned also in
Ref.~\cite{Bhattacharyya:2022ciw}.  To be specific, setting $\Phi uc$
couplings below $10^{-4}$ is enough to keep the contribution to
$D^0$--$\bar{D}^0$ mixing within its saturation limit. If $\Phi tc$
and $\Phi tu$ couplings are simultaneously present then they would
contribute to $D^0$--$\bar{D}^0$ mixing at one loop level, but it is
safe to assume that their products be less than $\sim 10^{-6}$.
Similarly, $\Phi bs$ couplings of size less than approximately $0.005$
would be enough to keep the tree level contribution to
$B_s$--$\bar{B_s}$ mixing within acceptable limit.  As regards the
leptonic Yukawa couplings, we keep only $\Phi \mu\tau$ as nonvanishing
to avoid stringent constraints on couplings involving electrons as
stated in Ref.~\cite{Bhattacharyya:2022ciw}.  The numerical hierarchy
of all these couplings is in line with the original motivation of
reproducing the quark masses and mixing, although actual numbers may
vary depending on the specific flavor models and the additional flavon
vevs. The spirit of the collider analysis undertaken in the present
paper is to take a simplified scenario with a few benchmark points by
keeping only those off-diagonal Yukawa couplings as sizable that
involve the top quark. A noteworthy point is that $S_3$-symmetric
flavor models require one of the flavors participating in Yukawa
interactions to be necessarily from the third generation, though the
off-diagonal Yukawa couplings involving the top quark are unrelated to
those involving the bottom quark
\cite{Bhattacharyya:2012ze,Bhattacharyya:2010hp}. With the above in
mind, the large production rate of the $t\bar t$ pair expected at the
high luminosity run of the LHC (HL-LHC) at 14 TeV gives us motivation
to carry out this top-specific exploration.

For simplicity, we assume $m_H > m_t$ and $m_\chi \sim (20-100)$ GeV
to focus on only one type of exotic state, namely $\chi$, being
produced on-shell from top quark decays. Admittedly, our collider
analysis is completely blind towards the CP nature of $\chi$.  To be
specific, we consider one of the pair produced top quarks to decay
into $W$ and $b$, and the other to $\chi$ and $j$, followed by $\chi$
decaying to $\mu$ and $\tau$.  Depending on the hadronic or leptonic
decay mode of $W^\pm$, we focus on two possible signal channels: (a)
$1b+3j+1 \mu + 1 \tau_h$ and (b) $ 1b + 1j + 1 \mu + 1\ell~
(\ell=e,\mu) + 1 \tau_h + \met$.  We consider only the hadronic decay
of $\tau$, denoted by $\tau_h$. We use sophisticated Machine Learning
techniques for multivariate analysis (MVA) to obtain maximally
enhanced signal significance.

The paper is structured as follows. In Section \ref{cons}, we show the
Feynman diagrams contributing to our signal events, discuss the
constraints on the relevant couplings and mention the benchmark points
selected for our studies.  Section \ref{analysis} contains exhaustive
collider analysis of the promising topologies using Machine Learning
techniques. We summarize and draw our conclusion in Section~\ref{Summary}.

\section{Selection of signal benchmark points} \label{cons} 

We probe light exotic spin-$0$ states coming from flavor violating top
quark decays. We rely on the huge production of $t\bar t$ events
during the HL-LHC run. We assume the CP-even state $H$ to be heavier
than the CP-odd state $\chi$ and fix $m_H$ to a value larger than
$m_t$, e.g. at 200 GeV. This is to avoid a 2-body decay of $t$ to $H$
and $j$. But since $\chi$ is assumed to be lighter than or equal to
100 GeV, $H$ (and $H^+$) should not be heavier than few hundred GeV to
respect the oblique parameter constraint \cite{Lu:2022bgw, Belanger:2024wca}, as stated
previously
\footnote{{Note that the simplified model parameter space we are
    studying, does not contain any $H^\pm$, but the inspirational
    flavor-specific mother models, or for that matter all multi-Higgs
    models, holistically do.  In order to remain conservative, we
    check consistency with $T$ and $S$ parameters, together with
    correlations among them, keeping the entire scalar multiplets in
    an extended scenario including the charged Higgs in hindsight,
    even though the latter does not feature in our specific model
    independent search domain.}}.
We have studied the correlation between the oblique parameters \cite{Lee:2012jn, Kanemura:2011sj,PhysRevD.106.035034} using the
scalar masses in the following ranges : $20 ~{\rm GeV} \leq m_\chi \leq
100 ~{\rm GeV}, ~ m_H = 200 ~{\rm GeV}, ~ 165 ~{\rm GeV} \leq m_{H^+}
\leq 250 ~{\rm GeV}$. Since the signals only contain the pseudoscalar $\chi$, we have the freedom to adjust $m_{H^+}$ to satisfy the constraints coming from the correlation of the oblique parameters. It could be
observed that for $m_\chi = $ 20 GeV, 60 GeV and 100 GeV, $m_H = 200 ~
\rm{GeV}$ and with proper choice of $m_{H^+}$ within the aforementioned range, the constraints coming from the oblique parameters are comfortably satisfied within $2 \sigma$ error bar of the
experimental limit ($S = 0.05 \pm 0.08,~ T = 0.09 \pm 0.07,~ \rho_{\rm ST} = 0.92.$) \cite{Lu:2022bgw, Belanger:2024wca}. A point to
note is that not only the masses but also the Yukawa couplings of $H$ (as well as of $H^+$) remain
essentially irrelevant for our studies.  Our primary focus is on the
decay channel $t \to \chi j$.

\begin{figure}[!h]
\centering
\subfloat[]{
  \label{fig:fd1}
        \centering
  \includegraphics[width=0.45\textwidth]{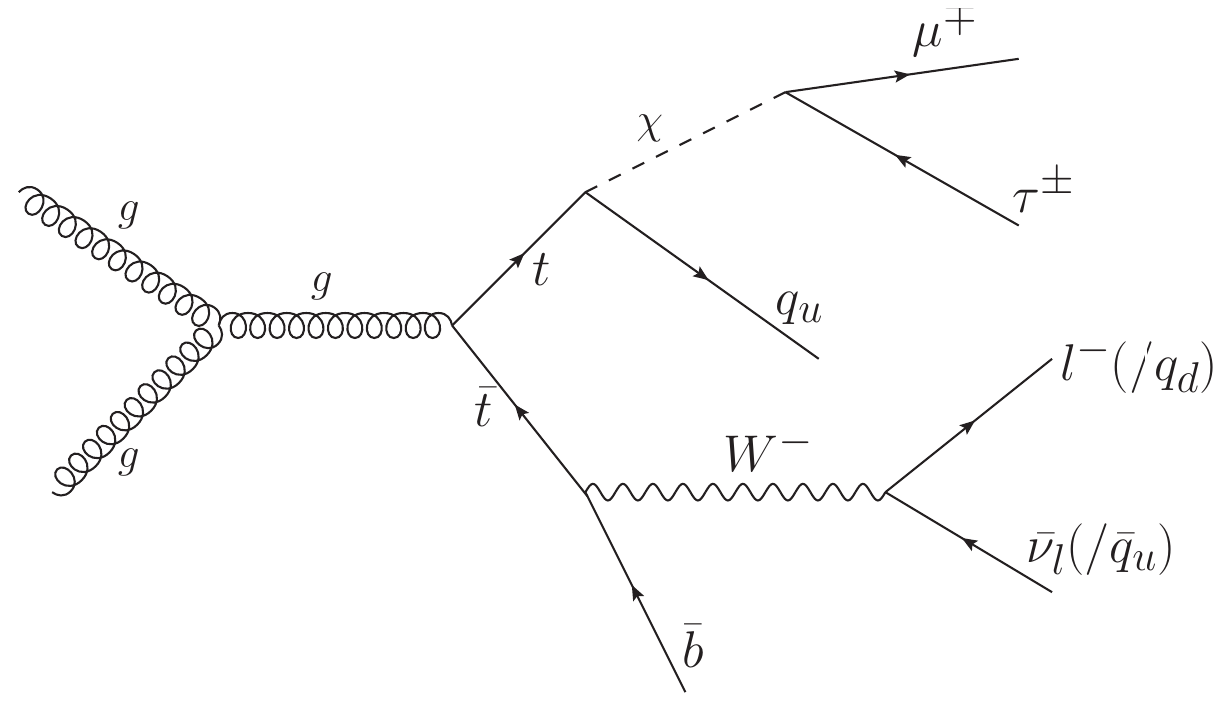}
}
\subfloat[]{
  \label{fig:fd2}
        \centering
  \includegraphics[width=0.40\textwidth]{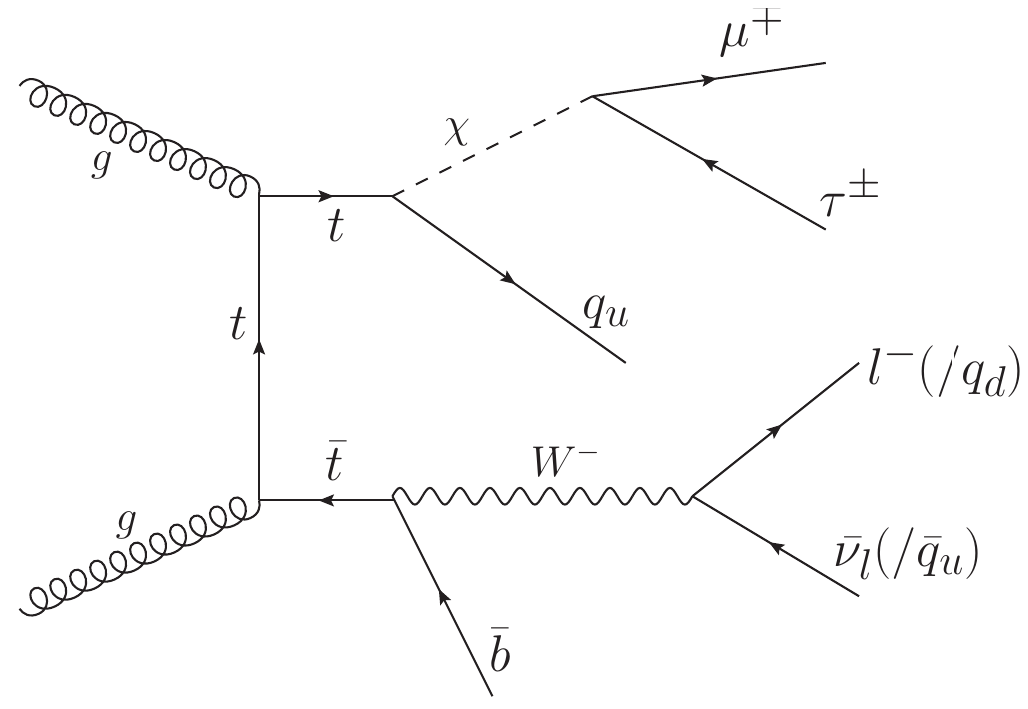}
}
\hspace{0.01\textwidth}
\subfloat[]{
  \label{fig:fd3}
        \centering
  \includegraphics[width=0.45\textwidth]{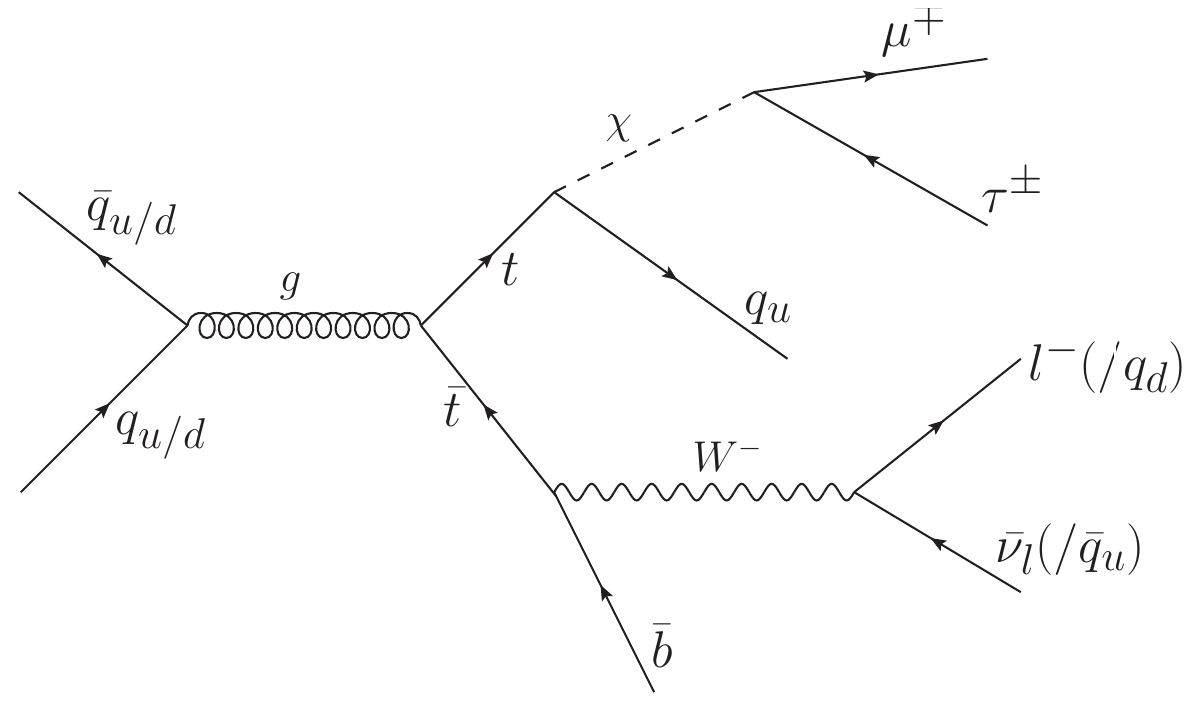}
}
\subfloat[]{
  \label{fig:fd4}
        \centering
  \includegraphics[width=0.40\textwidth]{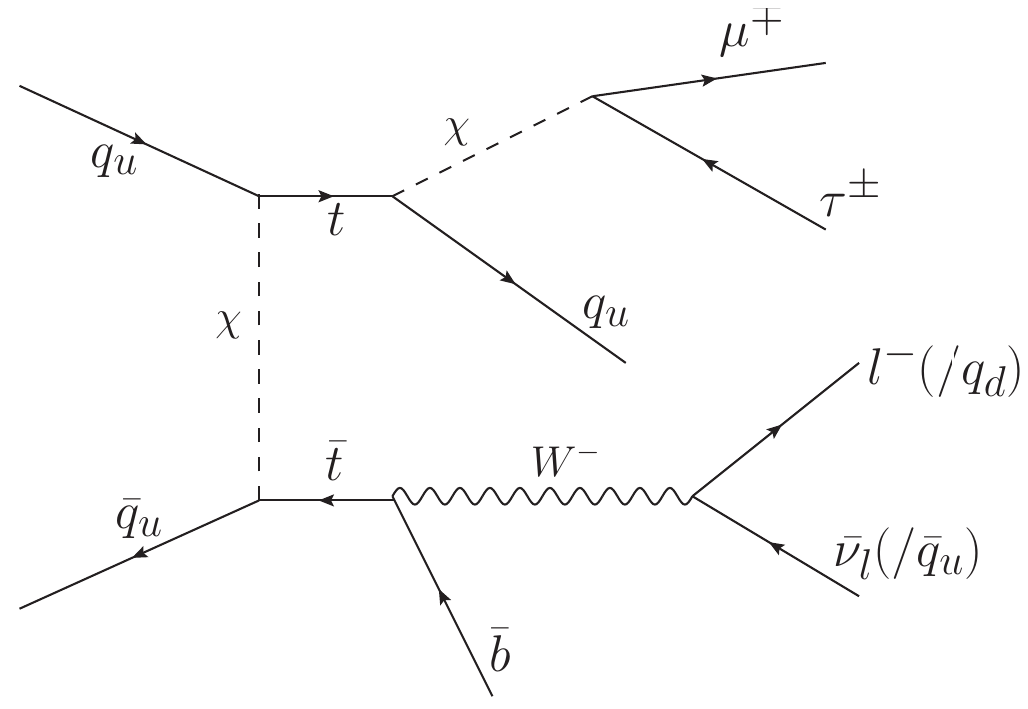}  
}
\caption{\small \em Feynman diagrams of the signal processes in the DL
  and SL channels.  Here, $q_u \equiv u,c$; $q_d \equiv d,s$.}
\label{fig:feynman1}
\end{figure}

Among the different possibilities regarding the signal final states,
we analyze only two specific channels involving leptons due to their
clean signatures. These are semi-leptonic (SL) and di-leptonic (DL)
channels, described below. We focus on one top quark decaying to $W$
boson and $b$-tagged jets, and the other top quark decaying to $\chi$
and a light jet. Subsequently, $W$ can decay to one lepton +
$\met$ or two light jets. To sum up, our final states are
\begin{itemize}
  \item $b+ 3j + \mu + \tau_{h}$ ~~ (SL),
  \item $b + j + \mu + \tau_{h} + \ell (=e,\mu) + \met$ ~~ (DL).
\end{itemize}
The relevant Feynman diagrams are shown in Figure \ref{fig:feynman1}.

What about the size of the off-diagonal Yukawa couplings of $\chi$? Is
there any guideline from the motivational models? The sizes of the
Yukawa couplings of $H$ are related to the entries of the CKM
matrix. But the Yukawa couplings of $\chi$ are not at all constrained
by that as there is no vev in the $\chi$ direction.  In
Ref.~\cite{Bhattacharyya:2012ze}, the $Hct$ coupling was taken to be
approximately $0.8$, by assuming that $H$ is heavier than the top
quark so that the latter does not have a 2-body decay into $H$ and
$c$. Since the $\chi$ is assumed to have a smaller mass than the top
quark in our analysis, our benchmark $\chi ct$ coupling has to be
significantly smaller.  The $\chi tc$ coupling is varied in the range
($0.001 - 0.01$).  This choice is consistent with the observation that
$t$ almost invariably decays to $b$ and $W$ with a branching fraction
of 0.998 as per the latest Particle Data Group \cite{Workman:2022ynf}.
If we turn on $\chi tu$ at the same time, $D^0$--$\bar{D}^0$ mixing is
triggered through one loop box graph forcing the product of $\chi tc$
and $\chi tu$ couplings to be at most ${10}^{-6}$. The choices are
consistent with the ATLAS and CMS searches for rare top
decays~\cite{TopQuarkWorkingGroup:2013hxj,
  ATLAS:2015iqc,CMS:2016uzc,ATLAS:2017beb,CMS:2017wcz,
  Workman:2022ynf}
\footnote{{These searches have set an upper limit on the Branching
    Ratios (BRs) of rare top decays like $t \to qh / qg/ qZ$, which
    vary roughly in the range ($10^{-2}$--$10^{-6}$) depending on the
    final states obtained by extrapolating the 7 TeV LHC data to 14
    TeV expectation with 3 ab$^{-1}$ luminosity. In our setup, since we are focusing on the
decay channel $t \to \chi j$ (here $j$ corresponds to light up-type quarks), those
    search limits are comfortably satisfied by our benchmark coupling
    range. }}.
In our scenario, $\chi$ decays to $\mu$ and $\tau$ with almost 100\%
branching ratio. As a result, though the exact choice of $\chi
\mu\tau$ coupling does not matter much, nevertheless we fix it at
$0.01$.  The effective signal cross section can be expressed as:
$\sigma_{\rm eff} = 2 \times \sigma (pp \rightarrow
t\bar{t}) \times{\rm Br}(t \rightarrow bW) \times{\rm Br}(t
\rightarrow c \chi) \times {\rm Br}(\chi \rightarrow \mu \tau) \times
            {\rm Br}(W \rightarrow \ell \nu / jj)$.
The overall factor 2 is a combinatoric factor arising from top
(anti-top) decay.  We summarize our
benchmark values in Table.~\ref{tab:BPs}
\footnote{{It is worth noting that our neutral exotic states ($\chi$
    and $H$) cannot be constrained by conventional LHC new scalar
    search strategies. The primary production mechanism of a neutral
    scalar at LHC relies on gluon fusion which proceeds through top
    quark loops. Vector boson fusion and associated productions also
    leave a sizable contribution to the production. In the absence of
    a diagonal Yukawa coupling, the gluon fusion mechanism is
    inoperative. On top of that, the absence of $HVV$ ($\chi VV$ being
    absent anyway) switches off the entire conventional mode of
    production for $\chi$ or $H$ search. This is why a standard
    package like \texttt{HiggsTools}~\cite{Bahl:2022igd} is
    inapplicable for constraining our benchmark points, thus
    attributing enhanced liberty to the choice of the latter.}}.
\begin{table}[h!]
  \begin{center}
    \begin{tabular}{| c | c | } \hline
      Parameter & Range \\
      \hline
      $m_{\chi}\,{\rm (GeV)}$ & 20, 60, 100     \\ \hline
      $Y^\chi_{\mu\tau}$        & 0.01            \\ \hline
      $Y^\chi_{ct}$             & 0.001 - 0.01    \\ \hline
    \end{tabular}
    \caption{\small \em Benchmark choices for masses and couplings.}
    \label{tab:BPs}
  \end{center} 
\end{table} 

\section{Collider Analysis} \label{analysis}

To perform the collider analysis, we first simulate the signal
benchmark points (Table \ref{tab:BPs}) and the samples of relevant SM
backgrounds (Table \ref{tab:baseline}). After applying a few
pre-selection conditions on both signal and background events in order
to select coarse signal regions, we proceed to perform MVA techniques.
Finally, we estimate the required integrated luminosity to achieve a
$5\sigma$ discovery and a $2\sigma$ exclusion. In the following
Subsections, we shall present the details of MVA, which are performed
to segregate the signal from corresponding backgrounds for each
benchmark point.

\subsection{Monte Carlo simulation of signal and background processes}

We start our analysis by implementing the Yuakawa Lagrangian in
\texttt{FeynRules} \cite{Alloul:2013bka} to generate Universal
FeynRules Output (UFO). In the next step, the UFO is interfaced with
the event generator to simulate the signal. Both the signal and
background events are generated using \texttt{MadGraph5\_aMC@NLO}
\cite{Alwall:2014hca}, providing the cross sections at the leading
order (LO). It is worth mentioning that in case of signal generation,
we have used a factorized approach, and hence ignoring all possible
interference effects and irreducible background. The \texttt{MadGraph}
syntax used for the semi-leptonic and di-leptonic event generation are
mentioned in the following:

\vspace*{0.5cm}

\noindent
{\em Semi-leptonic (SL) channel\,: }\\
{\texttt {\small {\footnotesize
define p = 21~ 2~ 4~ 1~ 3~ -2~ -4~ -1~ -3~ 5~ -5~ $\#$ pass to 5 flavors\\
define j = p\\
define qu = u~ u$\sim$~ c~ c$\sim$\\
define qd = d~ d$\sim$~ s~ s$\sim$~ b~ b$\sim$\\
define ln$\sim$ = mu- ~ ta- \\
define ln = mu{\small +}~  ta{\small +}\\
generate  p p > t t$\sim$,~(t > qu x,~ x > ln ln$\sim$),~(t$\sim$ > b$\sim$ w-, ~w- > qu qd)\\
add process p p > t t$\sim$,~(t > b w+, ~w+ > qu qd),~(t$\sim$ > qu x,~ x > ln ln$\sim$)\\
output S3\_Xmuta\_Wjj
}}}

\noindent
{\em Di-leptonic (DL) channel\,: }\\
{\texttt {\small {\footnotesize
define p = 21~ 2~ 4~ 1~ 3~ -2~ -4~ -1~ -3~ 5~ -5~ $\#$ pass to 5 flavors\\
define j = p\\
define qu = u~ u$\sim$~ c~ c$\sim$\\
define qd = d~ d$\sim$~ s~ s$\sim$~ b~ b$\sim$\\
define ln$\sim$ = mu- ~ ta- \\
define ln = mu{\small +}~  ta{\small +}\\
define ln1$\sim$ = e- ~mu-\\
define ln1 = e+ ~mu+\\
define vl1 = ve ~vm\\
define vl1$\sim$ = ve$\sim$ ~vm$\sim$\\
generate p p > t t$\sim$,~(t > qu x, ~x > ln ln$\sim$),~(t$\sim$ > b$\sim$ w-, ~w- > ln1$\sim$ vl1$\sim$)\\
add process p p > t t$\sim$,~(t > b w+, ~w+ > ln1 vl1),~(t$\sim$ > qu x, ~x > ln ln$\sim$)\\
output S3\_Xmuta\_Wlnu
}}}

\vspace*{0.5cm}
For the evaluation of both signal and background cross
sections, we employ \texttt{NN23LO1} as the parton distribution
function (PDF) \cite{NNPDF:2014otw}. The $\tau$ decays are simulated
within the \texttt{TAUOLA} package integrated in \texttt{PYTHIA-8}. These
parton level events are then passed through \texttt{PYTHIA-8}
\cite{Sjostrand:2014zea} for showering and hadronization. To
incorporate the detector effects, the resulting events are finally
processed through the fast detector simulation package
\texttt{Delphes-3.4.2} \cite{deFavereau:2013fsa} using the default CMS
card. Within Delphes, we use the anti-$k_T$ jet clustering algorithm
\cite{Cacciari:2008gp} using the \texttt{FastJet} package
\cite{Cacciari:2011ma}. The respective tagging efficiencies for the
$b$ and $\tau$-tagged jets have been parametrically incorporated
within the default CMS card.

Among all relevant backgrounds corresponding to two different signal
final states SL and DL, the most dominant is the $t\bar{t}$ pair
production. This background sample is generated by matching up to two
jets. Fully leptonic decays of $t\bar{t}$ pair introduce two leptons
in the final state while only one lepton emerges from the
semi-leptonic decay of $t\bar{t}$ pair. In addition, associated
production of single top with $W$ boson is another significant
contributor to the backgrounds. Sizable contributions also arise from
$t\bar{t}h$, $t\bar{t}V$, $VV$, and the QCD-QED $b\bar{b}Z^*/\gamma^*
\to b\bar{b}\tau^+ \tau^-$ processes tabulated in Table
\ref{tab:baseline}.

\subsection{Pre-selection criteria}

Relevant acceptance cuts on some of the kinematic variables are
required to be applied to identify different particles within the
finite size of the detectors.  For example, the transverse momentum of
each particle ($p_T$) should be above a particular threshold to
maintain optimum identification efficiency. Rejecting the particles
with low $p_T$ helps suppress the huge background contributions coming
from the QCD processes. Before performing an exhaustive collider
analysis, we first apply a set of acceptance cuts (C0) on some of the
pertinent kinematic variables as mentioned in
Table~\ref{tab:objSel}. Although, for signal and for a few background
processes (mentioned in Table \ref{tab:baseline}), all these cuts are
imposed at the generation level, for some background processes the
same cuts are also applied at the analysis level during object
selection to keep everything under the same roof. Next, we apply
a few more cuts on the lepton and jet multiplicity to achieve the same
final state as represented in Figure \ref{fig:feynman1}. Below, we
describe all the pre-selection cuts (C0 - C5) applied to the signal
and background events to select broad signal regions.

\begin{itemize}

\item [] {{\bf C0}\,:} The acceptance cuts consist of some basic
  selection criteria for leptons ($e, \mu, \tau$) and jets, imposed on
  the following set of kinematic variables: $(a)$ transverse momentum
  $p_T$, $(b)$ pseudo-rapidity $\eta$, and $(c)$ angular separation
  between $i$-th and $j$-th objects $\Delta R_{i,j}$ which is defined
  in terms of the azimuthal angular separation $(\Delta \phi_{ij})$
  and pseudo-rapidity difference $(\Delta \eta_{ij})$ between two
  objects $i$ and $j$ as $\sqrt{(\Delta \eta_{ij})^2 + (\Delta
    \phi_{ij})^2}$. The threshold values of these variables are quoted
  in Table~\ref{tab:objSel}.

  \begin{table}[!h]
    \begin{center}
      \footnotesize\setlength{\extrarowheight}{2pt}
      \begin{tabular}{|l|l|}
        \hline Objects & Selection cuts \\ \hline \texttt{$e$} &
        $p_{T} > 10$~{\rm GeV}, $~|\eta| < 2.5$ \\ \texttt{$\mu$} &
        $p_{T} > 10$~{\rm GeV}, $~|\eta| < 2.4$, $~\Delta R_{\mu e} >
        0.4$ \\ \texttt{$\tau_{h}$} & $p_{T} > 20$~{\rm GeV}, $~|\eta|
        < 2.4$, $~\Delta R_{\tau_h, e/\mu} > 0.4$
        \\ \texttt{$light\,jet$} & $p_{T} > 20$~{\rm GeV}, $~|\eta| <
        4.7$, $~\Delta R_{{\rm light\,jet}, e/\mu} > 0.4$
        \\ \texttt{$b\,jets$} & $p_{T} > 20$~{\rm GeV}, $~|\eta| <
        2.5$, $~\Delta R_{{\rm b\,jet}, e/\mu} > 0.4$ \\ \hline
      \end{tabular}
      \footnotesize
    \end{center}
    \caption{\small \em Summary of the acceptance cuts.}
    \label{tab:objSel}
  \end{table}

\item [] {{\bf C1}\,:} In both DL and SL channels, $\chi$ always
  decays to $\mu^{\pm}\tau^{\mp}$. But $W$ decays leptonically
  (hadronically) for the DL (SL) channel. We always identify $\tau$
  through its hadronic decay. Thus for the DL channel, final states
  each with at least one $\mu$ ($e \mu / \mu \mu$) is ensured, whereas
  for the SL channel, we demand exactly one $\mu$ and no $e$ in the
  final state.

\item [] {{\bf C2}\,:} In the final state we require exactly one
  $\tau$ jet, i.e. $\tau_h$, for both the SL and DL channels.

\item [] {{\bf C3}\,:} One of the pair produced top quarks in the
  signal decays to $b(\bar{b})W^-(W^+)$ and the other to $\chi$ and a
  light jet.  So, one $b$ tagged jet is required to be present in the
  final states of both channels. Apart from that, we also apply a cut
  on the number of light jets. For the DL channel, we demand at least
  one light jet in the final state, but for the SL analysis, because
  of the hadronic decay of $W$, the final state consists of minimum
  three light jets.

\item [] {{\bf C4}\,:} In the DL channel, the signal topology does not
  allow for a pair of opposite sign same flavor (OSSF) leptons in the
  final state arising out of the decay of $Z$-boson. Thus to exclude
  the $Z$-peak, we veto the events with a pair of OSSF leptons having
  an invariant mass $M_{\ell^{+}\ell^{-}}$ in the following window:
  $M_Z-10 ~ {\rm GeV}\,<\,M_{\ell^{+}\ell^{-}}\,<\,M_Z+10$ GeV.
\begin{table}[!h]
  \begin{center}
    {\footnotesize
    \begin{tabular}{|l|c|c|c|}
      \hline
      Process   & cross section (pb)    & \multicolumn{2}{c|}{Yields (${\cal L}\,=\,3\,{\rm ab^{-1}}$)} \\
      \hline\hline
      \multicolumn{4}{l}{\texttt{Signals}} \\
      \hline\hline
      \texttt{($m_\chi,\,Y^\chi_{ct}$)}    & \texttt{NLO ~$[{\rm DL\,||\,SL}]$}            &  \texttt{DL} & \texttt{SL}   \\
      \hline
      \texttt{$20,\,0.005$}                           & $0.008\,||\,0.026$                     &  $756$    & $3152.4$  \\
      \texttt{$20,\,0.01$}                            & $0.033\,||\,0.103$                     &  $2796$   & $11262$  \\
      \hline
      \texttt{$60,\,0.005$}                           & $0.012\,||\,0.036$                     &  $2372.4$   & $8696.4$ \\
      \texttt{$60,\,0.01$}                            & $0.048\,||\,0.144$                     &  $9559.2$   & $34670$ \\
      \hline
      \texttt{$100,\,0.005$}                          & $0.007\,||\,0.019$                     &  $1636.8$   & $5733.6$ \\
      \texttt{$100,\,0.01$}                           & $0.027\,||\,0.084$                     &  $6374$     & $23196.0$ \\
      \hline\hline
      \multicolumn{4}{l}{\texttt{SM Backgrounds}} \\
      \hline\hline
      \texttt{$t\bar{t}\,\to\,2\ell\,+\,jets$}           & $119.34$ [NNLO]\cite{WinNT}            & $1057603.13$      & $4314109.11$ \\
      \texttt{$t\bar{t}\,\to\,1\ell\,+\,jets$}           & $368.1$ [NNLO]\cite{WinNT}             & $7112.07$         & $7334375.95$ \\
      \texttt{$tW$}                                      & $34.81$ [LO]                           & $2463.11$         & $23907.39$ \\
      \texttt{$Z\,\to\,\tau^+\tau^-\,+\,jets$}           & $803$ [NLO]                            & $419.76$          & $20642.35$ \\
      \texttt{$t\bar{t}(W\,\to\,\ell\nu)\,+\,jets$}      & $0.25$ [NLO]                           & $2113.22$         & $8566.4$ \\
      \texttt{$t\bar{t}(W\,\to\,qq)$}                    & $0.103$ $^{(*)}$ [LO]                  & $207.74$          & $2410.91$ \\
      \texttt{$t\bar{t}(Z\,\to\,\ell^+\ell^-)\,+\,jets$} & $0.24$ [NLO]\cite{Kardos:2011na}       & $4228.31$         & $11606.32$ \\
      \texttt{$t\bar{t}(Z\,\to\,qq)$}                    & $0.206$ $^{(*)}$ [NLO]\cite{Kardos:2011na}   & $404.22$          & $4729.34$ \\
      \texttt{$WZ\,\to\,3\ell\nu\,+\,jets$}              & $2.27$ [NLO]\cite{Campbell:2011bn}     & $2404.83$         & $2688.63$ \\
      \texttt{$WZ\,\to\,2\ell\,2q$}                      & $4.504$ [NLO]\cite{Campbell:2011bn}    & $1275.99$         & $13659.67$ \\
      \texttt{$ZZ\,\to\,4\ell$}                          & $0.187$ [NLO]\cite{Campbell:2011bn}    & $169.42$          & $95.15$ \\
      \texttt{$t\bar{t}(h\,\to\,\tau^+\,\tau^-)$}        & $0.006$ $^{(*)}$ [LO]                  & $254.85$          & $661.17$ \\
      \texttt{$b\bar{b}\tau^+\tau^-$}                    & $0.114$ $^{(*)}$ [LO]                  & $36.89$           & $728.65$ \\
      \texttt{$WWW$}                                     & $0.236$ [NLO]                          & $62.0$            & $439.92$ \\
      \texttt{$WWZ$}                                     & $0.189$ [NLO]                          & $47.6$            & $510.03$ \\
      \texttt{$WZZ$}                                     & $0.064$ [NLO]                          & $22.48$           & $159.66$ \\
      \texttt{$ZZZ$}                                     & $0.016$ [NLO]                          & $3.42$            & $18.83$ \\
      \hline\hline\
      \texttt{${\rm Total~background}$}                  &                                        & $976565$          & $12692338$ \\ 
      \hline
    \end{tabular}
    }
  \end{center}
  \caption{\small \em Event yields at ${\cal L}\,=\,3\,\rm ab^{-1}$
    after applying the pre-selection criteria (C0 - C5) for the
    analyses in both the DL and SL decay channels. \\ ($*$): Some
    selections are applied at the generation (i.e. Madgraph)
    level. $p_T$ of jets\,(j) and $b$ quarks\,(b) $>\,20$ GeV, $p_T$
    of leptons\,($\ell$) $>\,10$ GeV, $|\eta|_{j/b}\,<\,5$,
    $|\eta|_\ell\,<\,2.5$ and $\Delta
    R_{jj/\ell\ell/j\ell/b\ell}\,>\,0.4$.  }
  \label{tab:baseline}
\end{table}

\item [] {{\bf C5}\,:} The next step is to select the leptons and the
  jets coming from $W$ boson in the DL and SL channels,
  respectively. As the decay of $\chi$ ensures the presence of one
  $\mu$, the two possible combinations of leptons in the final state
  are $\mu \mu$ or $\mu e$ where the second lepton originates from the
  decay of $W$ in DL channel. While the selection of the $\mu e$ final
  state is trivial, it is quite challenging to figure out whether the
  muon is coming from $\chi$ or from $W$.  We first check the
  distribution of $\Delta R$ between $\tau_h$ and $\mu$ for the $\mu
  e$ combination.  We observe that in most of the signal events,
  spatial separation between $\mu$ (coming from $\chi$) and $\tau_h$
  is smaller than that of $\tau_h$ and $e$ originating from
  $W$. Although this feature is most prominent for the benchmark with
  the lowest $m_\chi$, it persists for the other two benchmarks as
  well. For the $\mu\mu$ combination as well the same argument holds.
  First we choose the $\mu$ closer to the $\tau_h$. Then we check if
  the $\mu$ and the $\tau_h$ have opposite charges and if it is the
  case, we consider that $\mu$ to be the decay product of
  $\chi$. Conversely, if they have same charges, we do not reject the
  event just on that basis. We check if the $\tau_h$ and the distant
  $\mu$ have opposite charge. If so, we keep that second $\mu$ as the
  decay product of $\chi$ and the first $\mu$ as the one coming from
  $W$. This algorithm does not guarantee the selection of the perfect
  combination, still it makes the selection more efficient. To
  validate this method, the distributions of $\Delta R$ between
  $\mu-\tau_h$ for $\mu\mu$ and $\mu e$ final states in DL channel are
  shown in Figure \ref{fig:drs}.  The negligible difference between
  the distributions prove that this algorithm is exempted from any
  obvious selection bias.  The backgrounds in those plots are:
  $TT+$jets, $TT(V)+$jets and $VV(V)$. The $TT+$jets contain $t\bar{t}
  \to bW$ for SL and DL decay modes; the $T(T)V+$jets include $TV$ and
  $TTV$ processes, {\em e.g.}  $tW+$jets and $t\bar{t}h$; while the
  $VV(V)$ include the di and tri-boson processes.
  \begin{figure}[!h]
	\centering
	\captionsetup{justification=centering,margin=1cm}
	\subfloat[Channel : DL ($\mu e$)]{
	  \label{fig:DL20}
	  \centering
	  \includegraphics[width=0.45\textwidth]{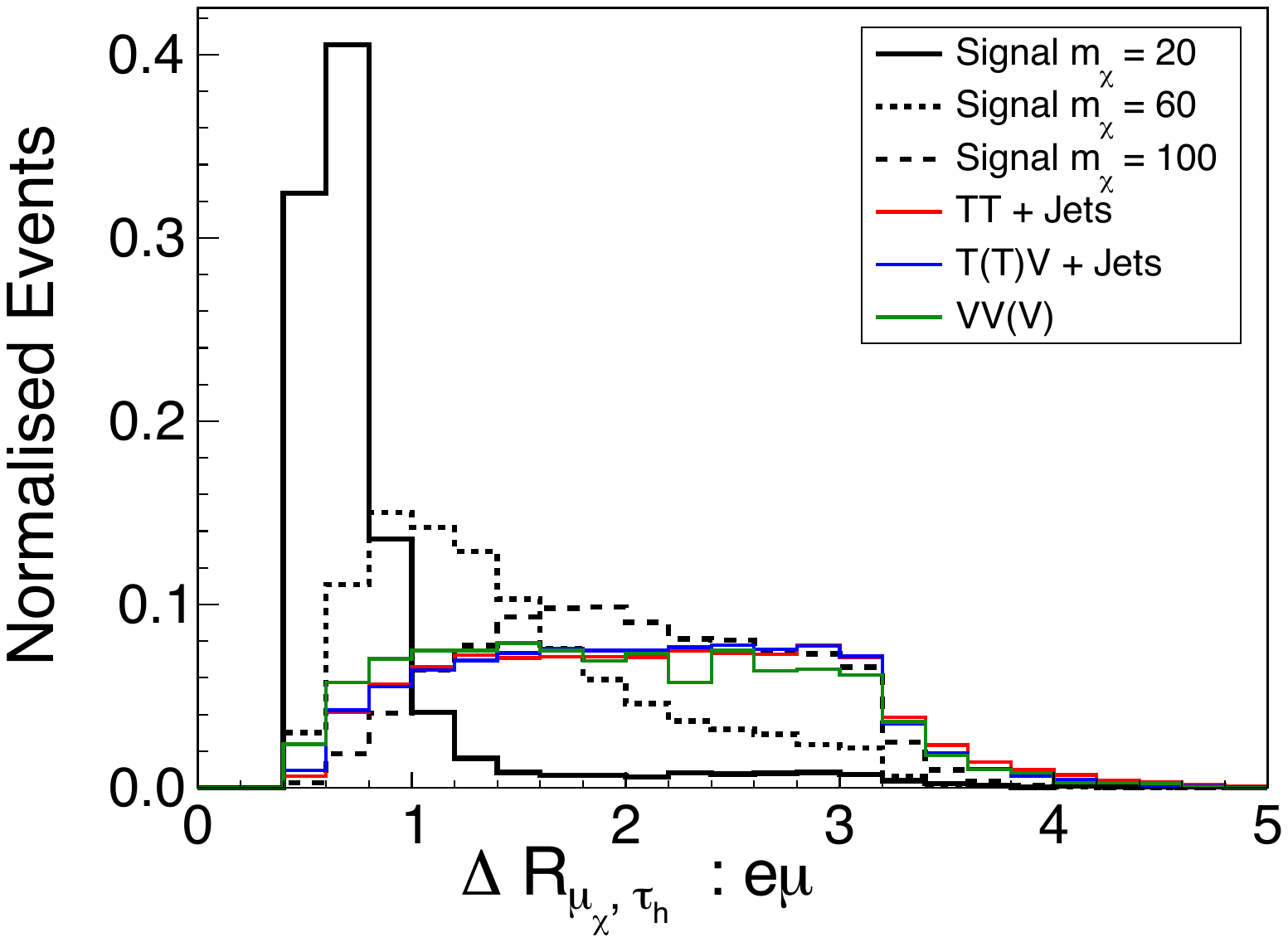}
	}
	\subfloat[Channel : DL ($\mu\mu$)]{
	  \label{fig:SL20}
	  \centering
	  \includegraphics[width=0.45\textwidth]{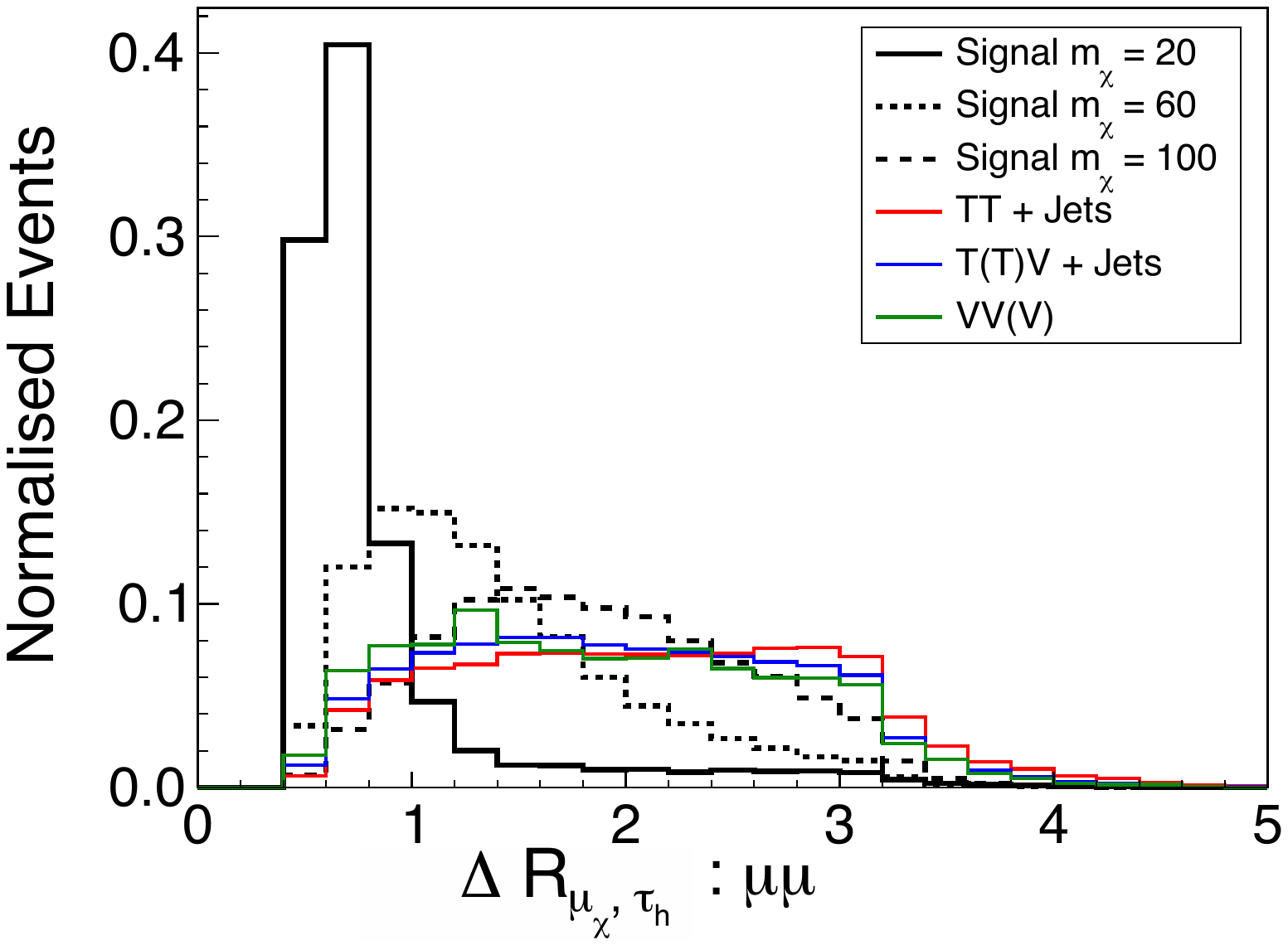}
	}
	\caption{\small \em $\Delta R$ between $\mu$-$\tau_h$ for (a) $e\mu$ (left) and (b) $\mu\mu$ (right) final states of DL channel. Note, $T \equiv t/\bar{t}, V \equiv W/Z$.}
	\label{fig:drs}
  \end{figure}
  The $\mu$ coming from $\chi$ is denoted by $\mu_\chi$ throughout 
  the rest of the discussion.

  ~ \, For the SL channel, we cannot differentiate whether the jets
  are coming from $t/\bar{t}$ or from $W$ at first sight. We first
  compute the invariant mass for each jet pair formed out of all three
  jets.  Then for each event we choose the jet pair having invariant
  mass closest to $M_W$ (within a $30$ GeV window around $M_W$), and
  tag the corresponding jet pair as originated from $W$. From the
  remaining jets, the leading one is selected as the light jet coming
  from $t/\bar{t}$. At the end, like in the DL channel, we demand that
  the $\mu_\chi$ and $\tau_h$ must have opposite charges.

\end{itemize}

After applying the selections mentioned above, we first reconstruct the mass of
$\chi$ for both channel. For the DL channel, there
are two sources of missing transverse energy $\met$, i.e. the
neutrinos generated in the hadronic decay of $\tau$ and those
generated from the leptonic decay of $W$. For SL channel, the single
source of $\met$ is the neutrinos generated from the hadronic decay of
$\tau$.  Using the collinear approximation \cite{ELLIS1988221} for
$\tau$ decay, we reconstruct the four momentum of $\tau$. The
fundamental assumption is that the decay products of $\tau$ are
boosted in the direction of $\tau$ itself since
$m_{\tau}\,\ll\,m_\chi$. Thus we take the projection of $~\met$ vector
along the direction of $\tau_h$ which estimates the $p_T$ of
$\nu_\tau$. Then we calculate the four momentum of the actual $\tau$
by modifying the transverse momentum ($p_T^{\tau_h}$) and energy
($E^{\tau_h}$) of the visible decay product $\tau_h$ by the factor
$\beta\,\equiv\,p_T^{\tau_h}/(p_T^{\tau_h}\,+\,p_T^\nu)$. The
transverse momentum and energy of the actual $\tau$ can be written as
:
\begin{equation} 
  p_T^\tau = \frac{p_T^{\tau_h}}{\beta}, ~ E^\tau = \frac{E^{\tau_h}}{\beta}
  \label{col}
\end{equation}
\begin{figure}[!h]
\centering
\subfloat[DL]{
  \label{fig:mcol_DL}
        \centering
  \includegraphics[width=0.45\textwidth]{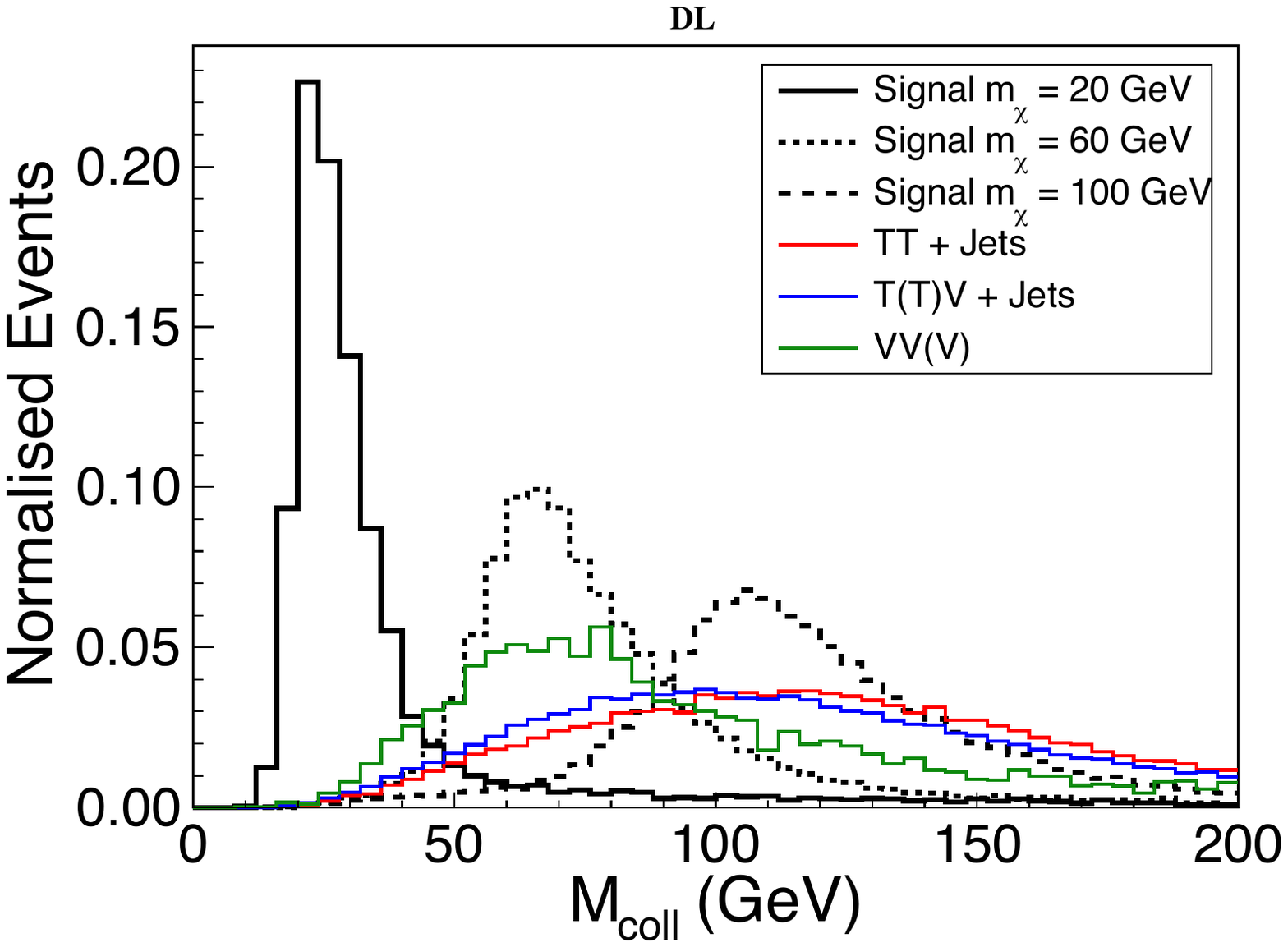}
}
\subfloat[SL]{
  \label{fig:mcol_SL}
        \centering
  \includegraphics[width=0.45\textwidth]{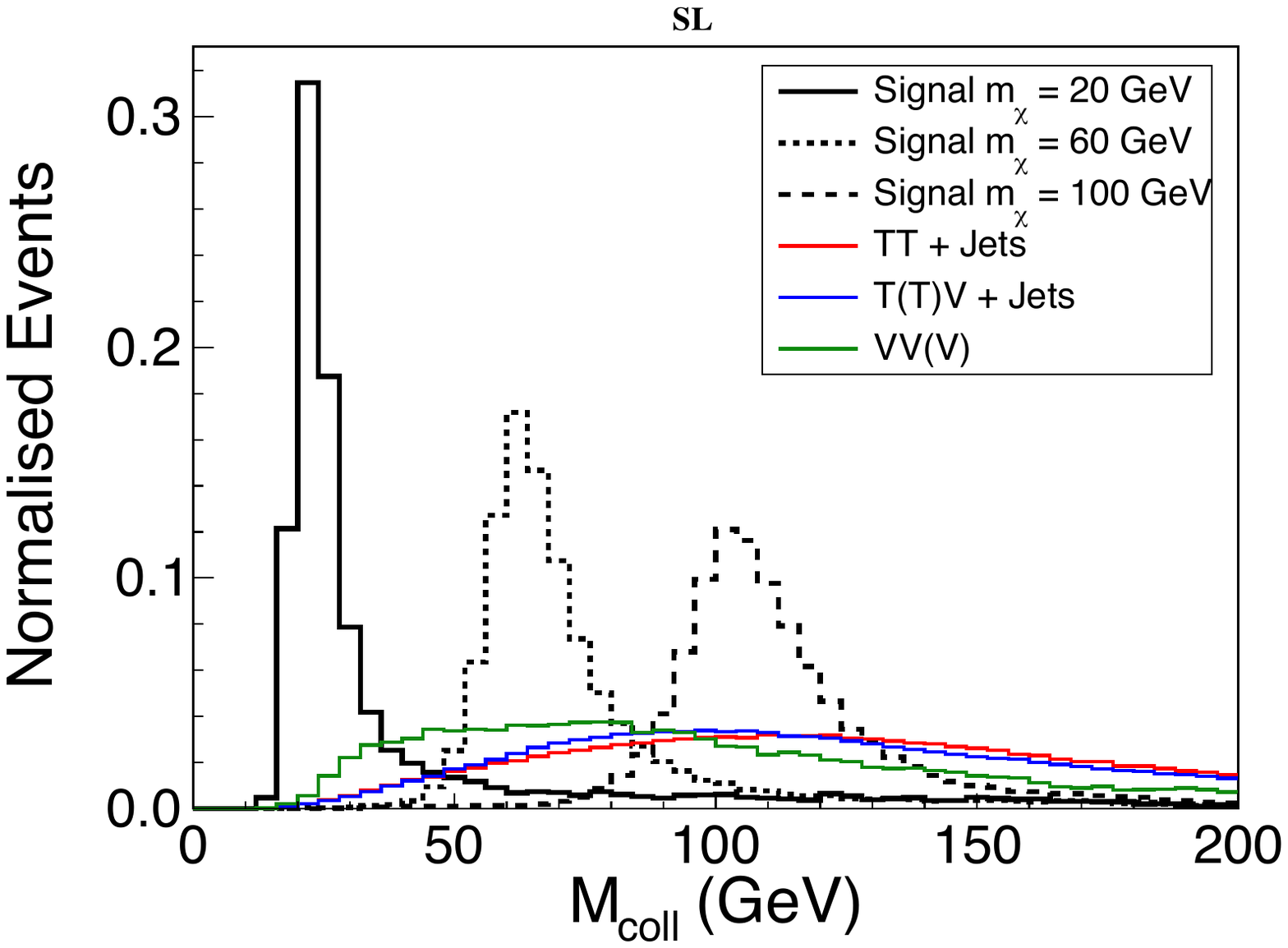}
}
\caption{\small \em Normalised distributions of collinear mass
  ($M_{coll}$) for the (a) DL (left) and (b) SL (right) channels.
Note, $T \equiv t/\bar{t}, V \equiv W/Z$.}
\label{fig:mcols}
\end{figure}
Although this approximation is not expected to yield the most accurate
result for the DL channel because of two different sources for
neutrinos, still we construct the collinear mass $M_{coll}$ of the
$\mu_\chi$-$\tau_h$ system using collinear approximation. In Figure
\ref{fig:mcols}, we depict the normalised distributions of $M_{coll}$
for both DL and SL channels. For SL channel, the distributions for
different benchmarks peak at $m_\chi = 20$, $60$ and $100$ GeV,
respectively. For the DL channel, the corresponding distributions peak
around $m_\chi$, and not exactly at $m_\chi$, for reasons stated
above.
 
 A huge number of background events (Table \ref{tab:baseline}) makes
 it quite challenging to classify the signal events efficiently from
 the SM backgrounds. Depending on the mass benchmarks,
 most of the distributions of various kinematic variables show overlapping 
 nature for signals and backgrounds. So, we avoid
 the traditional cut based method and deploy several MVA techniques to
 maximize the sensitivity of the analysis. The cross sections of the
 signals (scaled at next to NNLO+NNLL order by multiplying a $k$-factor
 of $1.8$\ \cite{WinNT}) and the SM background along with their yields
 after the pre-selection at an integrated luminosity
 $\mathcal{L}\,=\,3\,{\rm ab^{-1}}$ are tabulated in Table
 \ref{tab:baseline}. Next, we shall briefly describe three different
 MVA techniques that we employ to maximize the signal significance.

\subsection{Multivariate Analysis} \label{mva}

Mainly three different MVA techniques, namely, Decorrelated Boosted
Decision Tree (BDTD) \cite{Roe:2004na}, Extreme Gradient Boost
(XGBoost) \cite{Chen:2016:XST:2939672.2939785} and Deep Neural Network
(DNN) \cite{lecun2015deep} algorithms have been employed to estimate
the sensitivity of the signals over the SM backgrounds. For all the
three types of MVA techniques, the same set of input variables as
mentioned in Table \ref{tab:features} were used. Then we compare the
performances of the three methods and use the best one to estimate the
signal significance.

\begin{figure}[!h]
\centering \subfloat[DL]{
  \label{fig:DR_xlep_tauh_DL}
        \centering
  \includegraphics[width=0.45\textwidth]{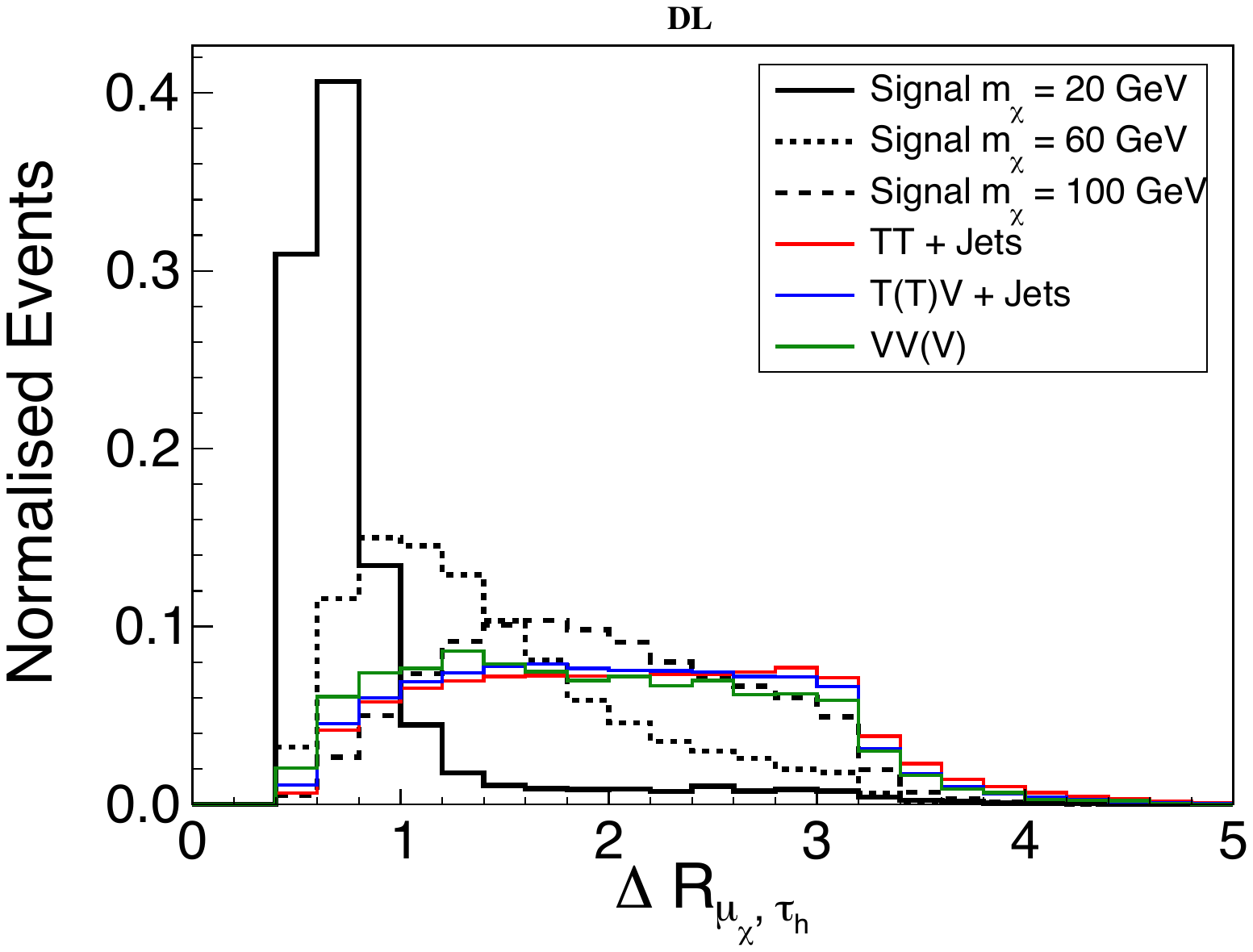}
}
\subfloat[SL]{
  \label{fig:DR_xlep_tauh_SL}
        \centering
  \includegraphics[width=0.45\textwidth]{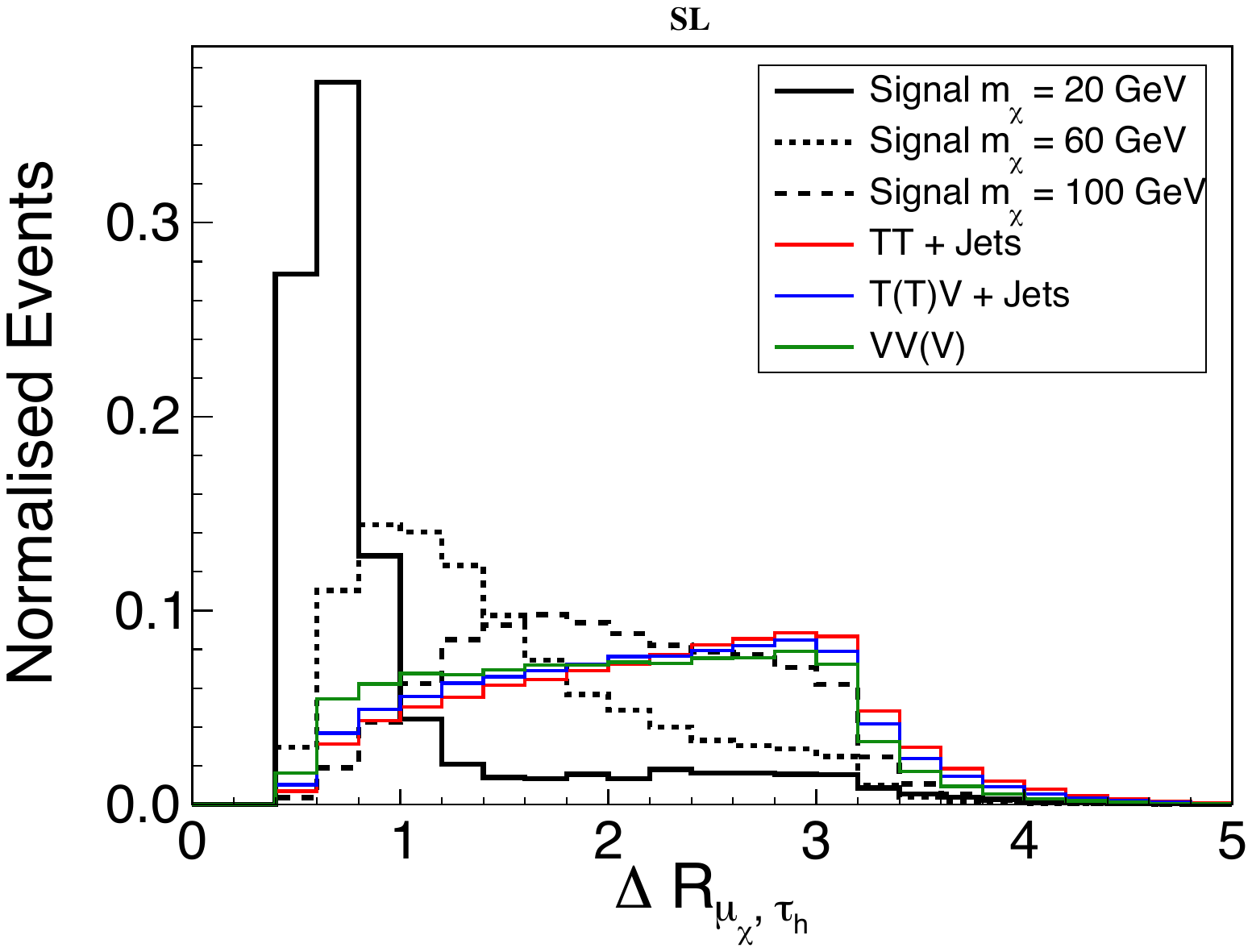}
}
\hspace{0.01\textwidth}
\subfloat[DL]{
  \label{fig:XlepPt_DL}
        \centering
  \includegraphics[width=0.45\textwidth]{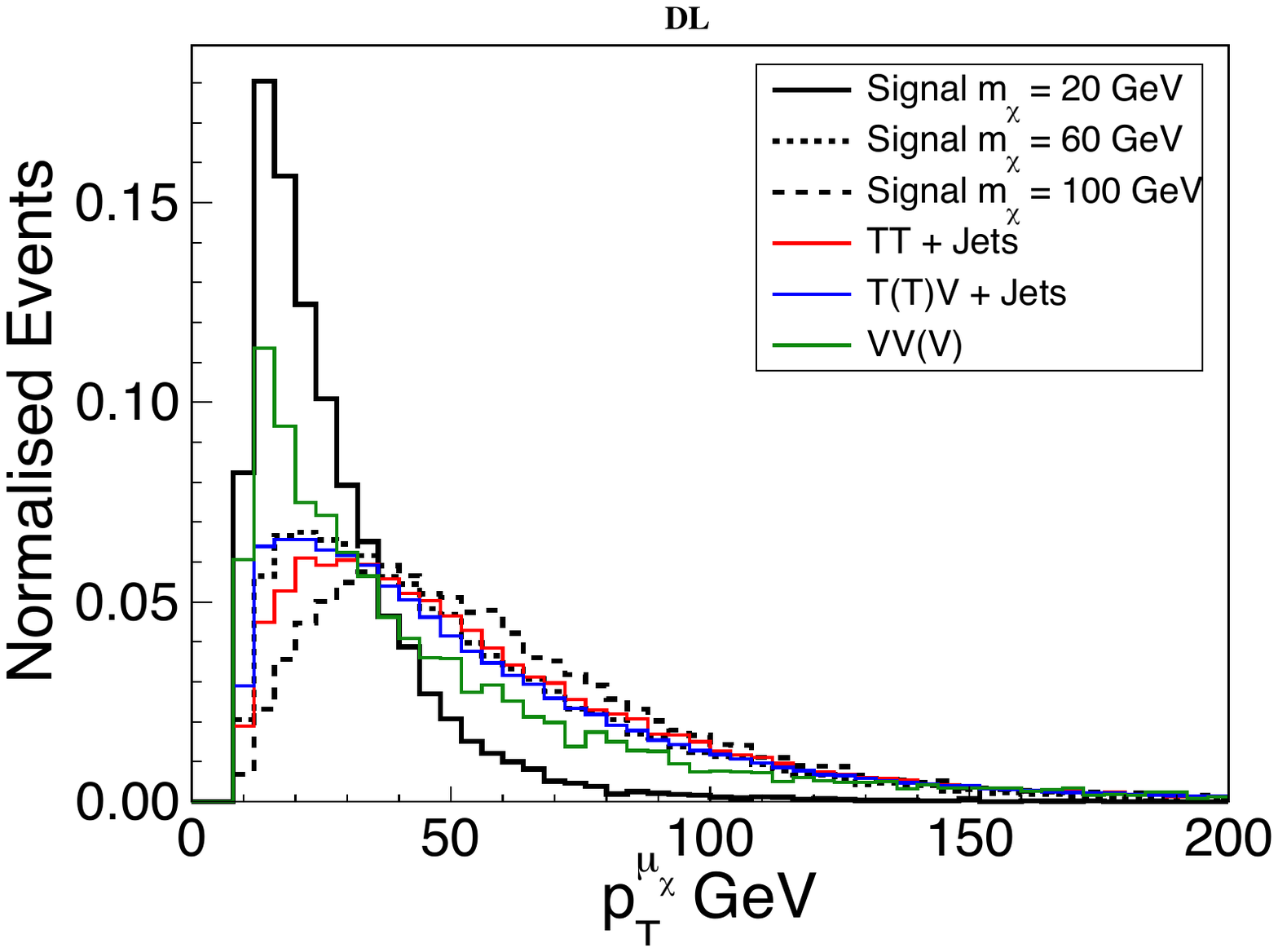}
}
\subfloat[SL]{
  \label{fig:XlepPt_SL}
        \centering
  \includegraphics[width=0.45\textwidth]{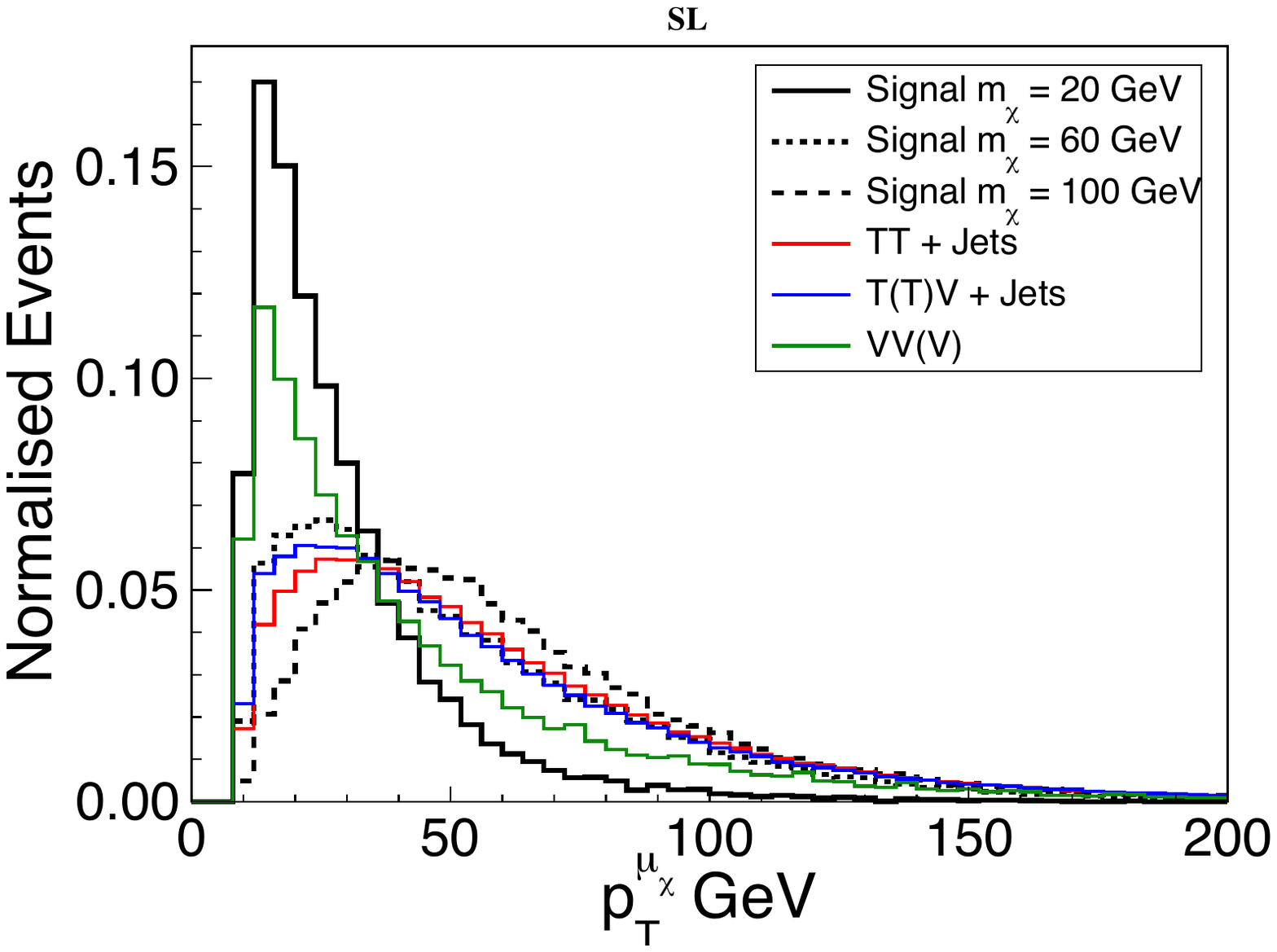}  
}
\hspace{0.01\textwidth}
\subfloat[DL]{
  \label{fig:smin_DL}
        \centering
  \includegraphics[width=0.45\textwidth]{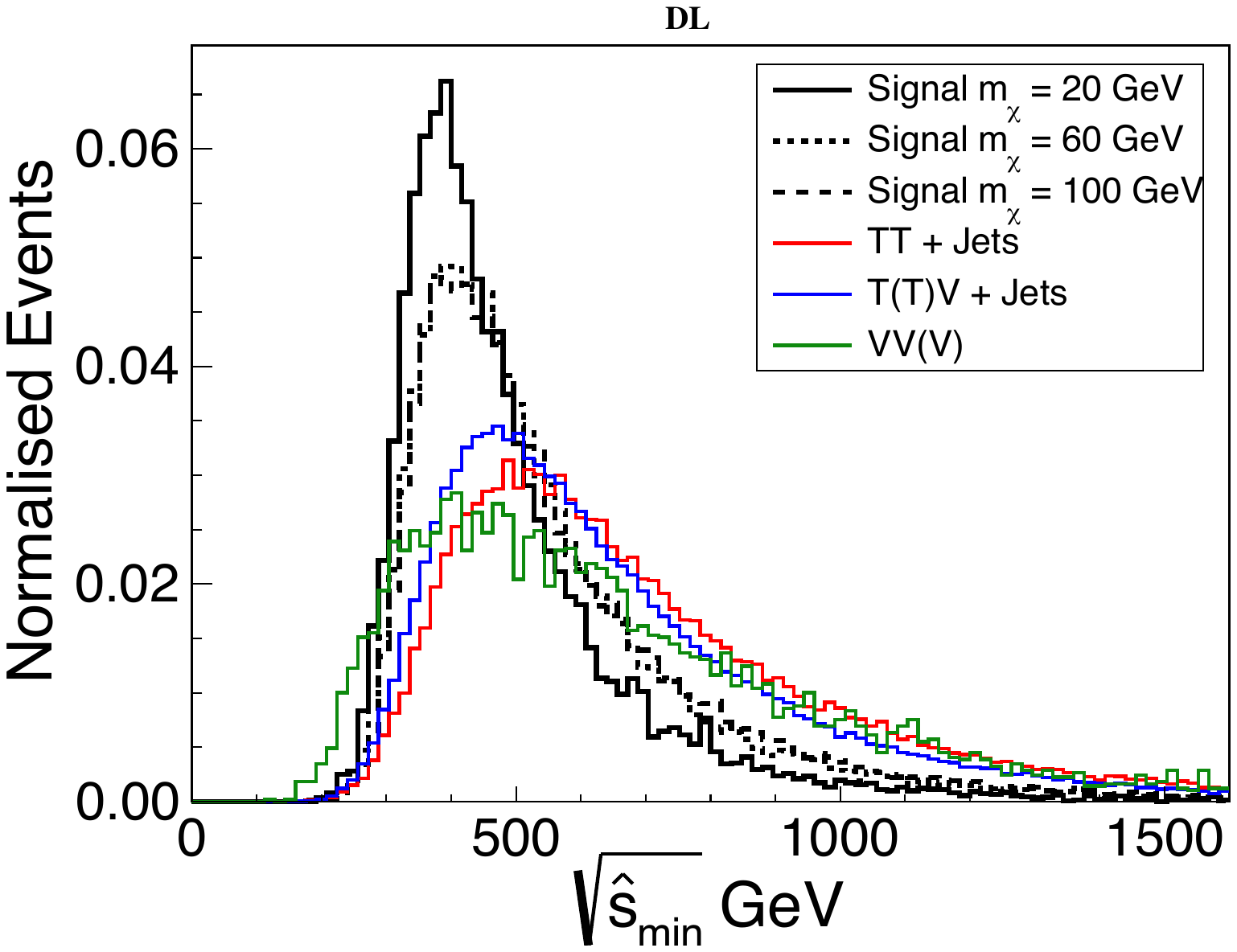}
}
\subfloat[SL]{
  \label{fig:smin_SL}
        \centering
  \includegraphics[width=0.45\textwidth]{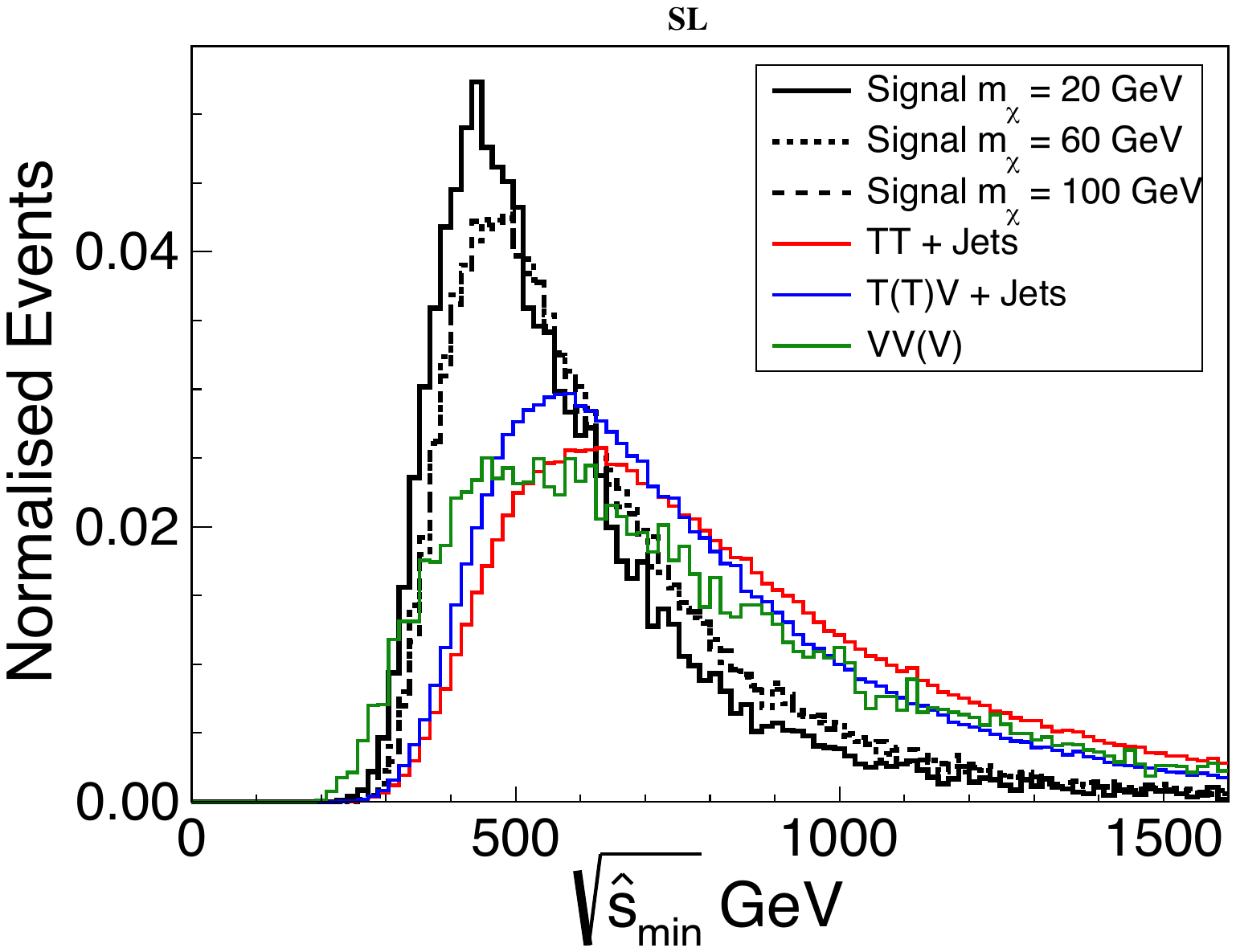}
}
\caption{\small \em Distributions of some kinematic variables. Column of figures 
  on the left side is for (a) DL channel and (b) SL channel plots are on the
  right column. The kinematic variables are: (a, b) $\Delta R$ between
  $\mu_\chi$ and $\tau_h$, (c, d) $p_T$ (in GeV) of $\mu_\chi$, and
  (e, f) $\sqrt{\hat{s}_{min}}$ {\it i.e.} minimum parton level center
  of mass energy for SL and DL channels. Note, $T \equiv t/\bar{t}, V \equiv W/Z$.}
\label{fig:features}
\end{figure}

Among all the input variables used for training the decision trees or
neural network, $\met$, $p_T$ of $\mu_\chi$ and $\tau_h$, $\Delta\phi$
and $\Delta R$ between different final state particles, and the scalar
sum of transverse momenta
\begingroup \setlength{\tabcolsep}{7pt}
\renewcommand{\arraystretch}{1}
\begin{table}[!h]
  \begin{center}
    \footnotesize\setlength{\extrarowheight}{2pt}
    \begin{tabular}{|l|c|c|c|}
      \hline
      Variables & Description & DL & SL  \\
      \hline
      \texttt{$p_T^{\tau_h},\,\eta^{\tau_h}$}        & $p_T$ and $\eta$ of $\tau_h$                                     &  \cmark & \cmark \\
      \texttt{$p_{T}^{bj},\,\eta^{bj}$}              & $p_T$ and $\eta$ of leading $b$ tagged jet                       &  \cmark & \cmark \\
      \texttt{$\met$}                                & Missing transverse energy                                        &  \cmark & \cmark \\
      \texttt{$p_{T}^{lj}\,\eta^{lj}$}               & $p_T$ and $\eta$ of light jet from $t/\bar{t}$                   &  \cmark & \cmark \\
      \texttt{$p_{T}^{Wj1},\,\eta^{Wj1}$}            & $p_T$ and $\eta$ of leading jet from $W$                         &  \xmark & \cmark \\
      \texttt{$p_{T}^{Wj2},\,\eta^{Wj2}$}            & $p_T$ and $\eta$ of sub leading jet from $W$                     &  \xmark & \cmark \\
      \texttt{$p_T^{\mu_\chi},\,\eta^{\mu_\chi}$}    & $p_T$ and $\eta$ of $\mu$ coming from $\chi$                     &  \cmark & \cmark \\
      \texttt{$p_T^{W_\ell},\,\eta^{W_\ell}$}        & $p_T$ of lepton coming from $W$                                  &  \cmark & \xmark \\
      \texttt{$\Delta R_{\mu_\chi, W_\ell}$}         & $\Delta R$ between leptons coming from $\chi$ and $W$            & \cmark & \xmark \\
      \texttt{$\Delta\phi_{\tau_h, W_\ell}$}         & $\Delta\phi$ between lepton coming from $W$ and $\tau_h$         & \cmark & \xmark \\
      \texttt{$\Delta\phi_{bj, W_\ell}$}             & $\Delta\phi$ between lepton coming from $W$ and lead $b$ jet     & \cmark & \xmark \\
      \texttt{$H_T$}                                 & Scalar sum $p_T$ of all jets                                     & \cmark & \cmark \\
      \texttt{$\Delta R_{\mu_\chi, \tau_h}$}         & $\Delta R$ between $\mu$ and $\tau_h$ coming from $\chi$         & \cmark & \cmark \\
      \texttt{$m_T^{\met, W_\ell}$}                  & $m_T$ of lepton from $W$ and $\met$                              & \cmark & \xmark \\
      \texttt{$\Delta\phi_{\,\met, \mu_\chi}$}       & $\Delta\phi$ between $\met$ and the $\mu$ from $\chi$            & \cmark & \cmark \\
      \texttt{$\Delta\phi_{\,\met, W_\ell}$}         & $\Delta\phi$ between $\met$ and the lepton from $W$              & \cmark & \xmark \\
      \texttt{$\Delta\phi_{\,\met, bj}$}             & $\Delta\phi$ between $\met$ and leading $b$ jet                  & \xmark & \cmark \\
      \texttt{$\Delta\phi_{\,\met, \tau_h}$}         & $\Delta\phi$ between $\met$ and $\tau_h$                         & \xmark & \cmark \\
      \texttt{$\Delta R_{Wj1, Wj2}$}                 & $\Delta R$ between the two jets from $W$                         & \xmark & \cmark \\
      \texttt{$m^{inv}_{\mu, \tau, t_{\ell j}}$}     & Invariant mass of reconstructed $\tau$, $\mu$ and light jet from $t/\bar{t}$     & \xmark & \cmark \\ 
      \texttt{$\sum\,p_T^{all}$}                     & scalar sum $p_T$ of all the final state particles                & \xmark & \cmark \\
      \texttt{$\Delta\phi_{\,lj, bj}$}               & $\Delta\phi$ between leading $b$ jet and leading light jet       & \cmark & \cmark \\
      \texttt{$\Delta R_{\tau_h, lj}$}               & $\Delta R$ between $\tau_h$ and leading light jet                & \cmark & \cmark \\
      \texttt{$\sqrt{\hat{s}_{min}}$}                & Minimum parton level center-of-mass energy                       & \cmark & \cmark \\
      \texttt{$cos\theta^*_{\tau_h, \mu_\chi}$}      & Cosine of the angle between $\chi$ and one of its                &        &       \\
                                                     & decay products in the rest frame of $\chi$                       & \cmark & \cmark \\
      \hline
    \end{tabular}
    \footnotesize
  \end{center}
  \caption{\em Input variables used(\cmark) or not(\xmark) for all of
    three MVA methods.}
  \label{tab:features}
\end{table}
of jets $(H_T)$ turn out to be the most important
ones\footnote{Importance of variables is checked by Toolkit for
Multivariate Data Analysis (TMVA) variable ranking for BDTD. For DNN
and XGBoost, we use permutation method using F-Score
\cite{Breiman2001}.}.  Figure \ref{fig:features} shows the normalised
distributions of a few important kinematic variables for all three
signal mass points and the backgrounds. Figures
\ref{fig:DR_xlep_tauh_DL} and \ref{fig:DR_xlep_tauh_SL} show the
normalised distributions of $\Delta R$ between the $\tau_h$ and $\mu$
coming from $\chi$ for DL and SL channels, respectively. The
distributions look similar for both the channels as expected. For
$m_\chi\,=\,20$ GeV, the decay products of $\chi$ are supposed to be
more boosted and hence collimated than the other two benchmark
points. That is why the distributions with lower $m_\chi$ peak at a
lower value of $\Delta R$.  Next we present the normalised
distributions of another important variable $p_T$ of $\mu$ coming from
$\chi$ for both decay channels in Figure \ref{fig:XlepPt_DL} and
\ref{fig:XlepPt_SL}. The distributions peak towards higher values of
$p_T$ for higher values of $m_\chi$ as expected.  We now mention
another important variable: minimum parton level
centre-of-mass energy i.e. $\sqrt{\hat{s}_{min}}$
\cite{Konar:2008ei} as shown in Figure \ref{fig:smin_DL} and \ref{fig:smin_SL}. This is a global inclusive variable for
determining the mass scale of any new physics in the presence of
missing energy in the final states.  This variable shows a good 
discriminating power as well to identify signals over backgrounds 
for both the DL and SL channels. In order to minimize any direct 
dependence on benchmark points, we refrain from utilizing any 
kinematic variable related to the reconstructed mass of the 
hypothetical particle that is yet to be seen. 

We conduct two-class classification tasks using all three MVA techniques. 
The "background" class comprises all SM background processes except $VVV$, 
while the "signal" class includes signal processes for all considered 
couplings per mass benchmark. $75\%$ of the total signal 
and background events are utilized for training. Following this, we evaluate 
the performances of the respective models using the entire dataset as 
prior testing on a subset (remaining $25\% $) results in negligible 
over-training. Additionally, $20\%$ events from the training set used as a 
validation set to monitor training performance. As an example, the ratio of 
the signal (for $m_\chi = 20$ GeV) : background
events for SL analysis is around $150000 : 680000$ and for DL it is
around $125000 : 160000$. While combining all the individual
background process, we kept a "sample-weight" {\it i.e.}
cross-section/total-events to maintain the ratio of different
backgrounds as they should appear in the real data. For signal, we
fixed it to 1. At the time of training in \texttt{Keras}, we used
equal class weights for these signal and background classes on top of
using sample-weight, to give equal importance to signal and background
classes.

The entire algorithm of BDTD is executed within the TMVA framework
\cite{Hocker:2007ht}. Here, {\it Adaptive Boost} \cite{FREUND1997119}
plays a crucial role for robust and efficient classification. To
achieve an optimum performance for each benchmark of both the DL and
SL channels, we adjust the parameters of BDTD as described in Table
\ref{tab:bdtdparams}.
\begin{table}[!h]
  \begin{center}
    \footnotesize\setlength{\extrarowheight}{1pt}     
    \begin{tabular}{|l|c|c|c|c|c|}
      \hline
      \multirow{2}{*}{Parameters} & \multirow{2}{*}{Description}                  & \multicolumn{3}{c|}{Values/Choices}    \\ \cline{3-5}
                                  &                                               & $m_\chi=20$ GeV & $m_\chi=60$ GeV & $m_\chi=100$ GeV\\
      \hline
      \texttt{n\_trees}           & Number of tress                               & $250$       & $250$       & $250$        \\  
      \texttt{max\_depth}         & Maximum depth of a Decision Tree              & $2$         & $2$         & $2$          \\
      \texttt{boost}              & Boosting mechanism for training               & AdaBoost    & AdaBoost    & AdaBoost     \\
      \texttt{n\_cuts : SL/DL}    & Number of iteration to find the best split    & $50/50$     & $46/31$     & $45/40$      \\
      \texttt{min\_node\_size}    & Minimum events at each final leaf             & $2.5\%$     & $2.5\%$     & $2.5\%$      \\  
      \hline
    \end{tabular}
    \footnotesize
  \end{center}
  \caption{\small \em BDTD parameters used for three different mass benchmark points.}
  \label{tab:bdtdparams}
\end{table}
XGBoost is another tree based method like BDTD with some additional
features. Unlike BDTD, it uses {\it Gradient Boost} for
classification.  To reduce over-training, some additional parameters
are used for pruning a decision tree and regularizing the cost
function defined as the difference between the true and predicted
output -- for details on XGBoost see
\cite{Chen:2016:XST:2939672.2939785}. Table \ref{tab:xgbparams} shows
the set of XGBoost parameters used for training the signal and
background samples.
\begin{table}[!h]
  \begin{center}
    \footnotesize\setlength{\extrarowheight}{1pt}     
    \begin{tabular}{|l|c|c|}
      \hline
      Parameters             & Description                             & Values/Choices    \\
      \hline
      \texttt{booster}       & Tree based learner               & $gbtree$ \\  
      \texttt{n\_estimators} & Number of decision trees           & $auto$   \\
      \texttt{max\_depth}    & Maximum depth of a Decision Tree                   & $3$      \\
      \texttt{$\eta$}        & Learning rate                           & $0.01$   \\
      \texttt{$\lambda$}     & Regularization parameter                & $0.01$   \\
      \hline
    \end{tabular}
    \footnotesize
  \end{center}
  \caption{\small \em  Details of XGBoost parameters.}
  \label{tab:xgbparams}
\end{table}
The third MVA technique that we have employed is Deep Neural Network
(DNN).  Unlike decision trees, DNN brings in several hidden layers
with multiple nodes.  Nonlinear activation functions at the nodes help
draw nonlinear boundary on the plane of the DNN variables to separate
signal from background. The complete DNN training has been performed
using {\tt keras} module of {\tt tensorflow-2.3.0}
\cite{tensorflow2015-whitepaper}. All the parameters used for DNN are
in Table \ref{tab:dnnparams}.
\begin{table}[htpb]
  \begin{center}
    \footnotesize\setlength{\extrarowheight}{1pt}     
    \begin{tabular}{|l|c|c|}
      \hline
      Parameters                 & Description                                   & Values/Choices  \\
      \hline
      \texttt{n\_hidden layers}  & Number of hidden layers                       & $5$    \\  
      \texttt{n\_nodes}          & Number of neurons in hidden layers            & $512$, $256$, $128$, $54$, $8$  \\
      \texttt{activation\_func}  & Function to modify outputs of every nodes     & $LeakyRelu$                     \\
      \texttt{loss\_function}    & Function to be minimised to get optimum model parameters      & $binary\_crossentropy$ \\
      \texttt{optimiser}         & Perform gradient descent and back propagation & $Adam$ \cite{Kingma:2014vow} \\  
      \texttt{eta}               & Learning rate                                 & $0.001$ \\
      \texttt{batch\_len}        & Number of events in each mini batch           & $3000$  \\ 
      \texttt{batch\_norm}       & Normalisation of activation output            & $True$  \\
      \texttt{dropout}           & Fraction of random drop in number of nodes    & $20\%$  \\
      \texttt{L2-Regularizer}    & Regularize loss to prevent overfitting        & $0.001$ \\      
      \hline
    \end{tabular}
    \footnotesize
  \end{center}
  \caption{\small \em Details of DNN parameters.}
  \label{tab:dnnparams}
\end{table}

\begin{figure}[!h]
\centering
\subfloat[DL ($m_\chi~=~60$ GeV)]{
  \label{fig:ROC_DL}
        \centering
  \includegraphics[width=0.45\textwidth]{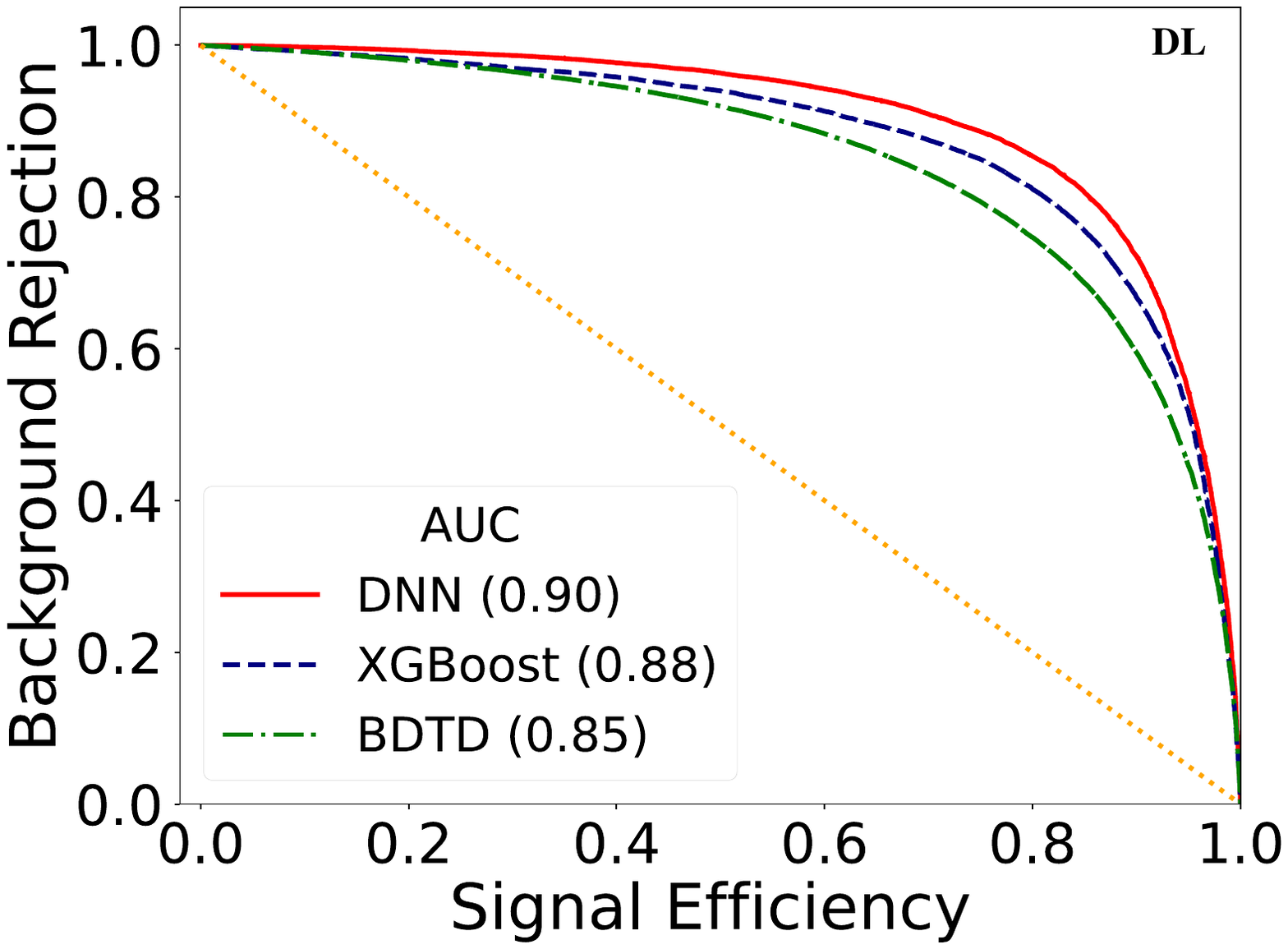}
}
\subfloat[SL ($m_\chi~=~60$ GeV)]{
  \label{fig:ROC_SL}
        \centering
  \includegraphics[width=0.45\textwidth]{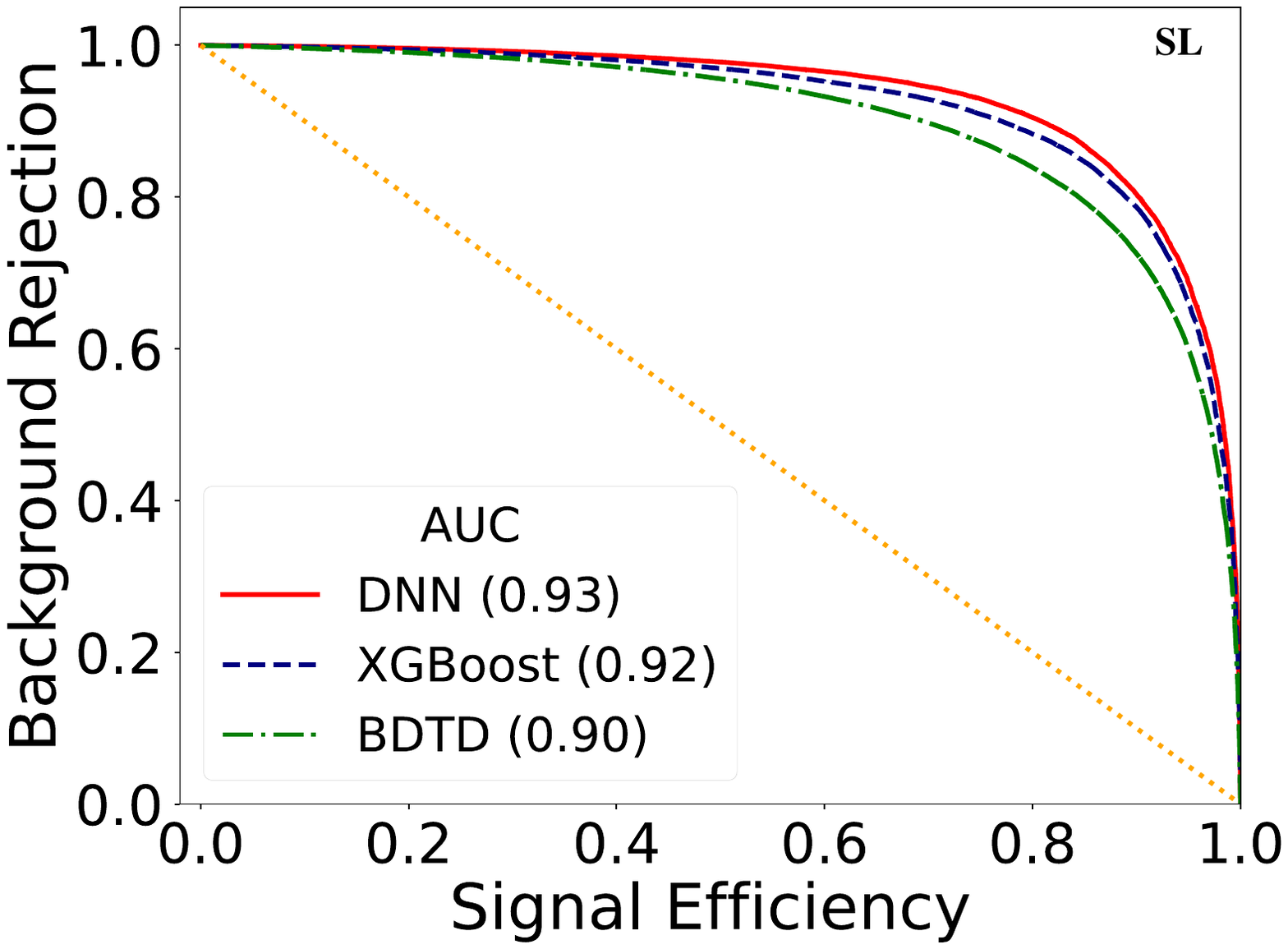}
}
\caption{\small \em ROCs of three MVA techniques for $m_\chi\,=\,60$ GeV.}
\label{fig:ROCs}
\end{figure}

All the tree based methods and neural network used in this analysis
rely on quite simple architectures. DNN was found to take a bit longer
time to finish than the other two methods. Each epoch for DNN was
observed to take around 30-45 seconds in a Tesla P100 GPU, while each
estimator of XGBoost needed around 10-15 seconds.  These three MVA
techniques discussed above deliver more or less similar
performances. In order to compare their responses, we plot the
receiver operating characteristic (ROC) curves for all the three
methods and compute the area under the curve (AUC) of each ROC in
Figure \ref{fig:ROCs}.  The degree of performance of the MVA
techniques increases with increasing AUC. The comparison shows that
the DNN performs somewhat better than BDTD and XGBoost in both the DL
and SL channels as reflected in Figure \ref{fig:ROCs} for
$m_\chi\,=\,60$ GeV. The other two choices of $m_\chi$ exhibit similar
behavior. We must admit that no conventional method for hyperparameter
tuning has been tried so far for executing these MVA techniques. A
proper tuning may change the degree of performances.  However, based
on the ROC curves obtained here, we shall elaborate the final results
only for the DNN technique.

\subsection{Results} \label{result}
For evaluation of the signal significance using DNN, we apply suitable
cuts on the respective DNN responses that maximize the
significance\footnote{For $B >> S$, the median significance can be
  simplified as ${\cal S} = \frac{S}{\sqrt{B}}$ \cite{Cowan:2010js},
  where $S$ and $B$ are the number of expected signal and background
  events, respectively. This expression does not take into account
  systematic uncertainties}. Then we try to find the required
luminosity to achieve a $2\sigma$ exclusion and $5\sigma$ discovery by
scanning over the coupling $Y^\chi_{ct}$. We also estimate the significance after
introducing a $5\%$ systematic uncertainty\footnote{In the presence of
  systematic uncertainty the modified significance is : ${\cal
    S}\,=\,\frac{S}{\sqrt{B\,+\,(\theta \times B)^2}}$,
  where $\theta$ is the uncertainty (in $\%$) on the number of
  background events.}  in total number of background events.
\begin{table}[htpb]
  \begin{center}
    \footnotesize\setlength{\extrarowheight}{1pt}     
    \begin{tabular}{|l|c|c|c|c|c|}
      \hline
      \multirow{3}{*}{Classifiers}& \multicolumn{4}{c|}{Signal efficiency\,($\%$)} & \multirow{3}{*}{Maximum significance}     \\ \cline{2-5}
                                  &     $95$    &    $80$   &    $65$   &   $50$   &                                           \\ \cline{2-5}
                                  & \multicolumn{4}{c|}{Background rejection\,($\%$)}  & ${\cal L} = 3\,{\rm ab^{-1}}$         \\ \cline{2-5}
      \hline
      BDTD                        &     $49$    &    $78$   &    $88$   &   $93$   &      $9.34$                               \\
      XGBoost                     &     $59$    &    $86$   &    $92$   &   $95$   &      $11.1$                               \\
      DNN                         &     $64$    &    $89$   &    $95$   &   $97$   &      $14.9$                               \\
      \hline
    \end{tabular}
    \footnotesize
  \end{center}
  \caption{\small \em A quantitative comparison of the three MVA
    techniques for the DL channel with $m_\chi\,=\,60$ GeV and
    $Y^\chi_{ct}\,=\,0.01$. }
  \label{tab:bestmva}
\end{table}
The benchmark point with $m_\chi = 20$ GeV still shows a good sensitivity
after incorporating a $5\%$ systematic uncertainty in the
background. The corresponding sensitivity significantly drops for
higher $m_\chi$ benchmark points -- see Figure \ref{fig:contours}.
Lighter the choice of $\chi$ the more pronounced is the difference in
kinematics of the signal compared to the background -- a feature
observed for all the three MVA techniques.

\begin{figure}[!h]
  \centering
  \captionsetup{justification=centering,margin=1cm}
  \subfloat[DL : $m_\chi\,=\,20$ GeV]{
    \label{fig:DL20}
    \centering
    \includegraphics[width=0.45\textwidth]{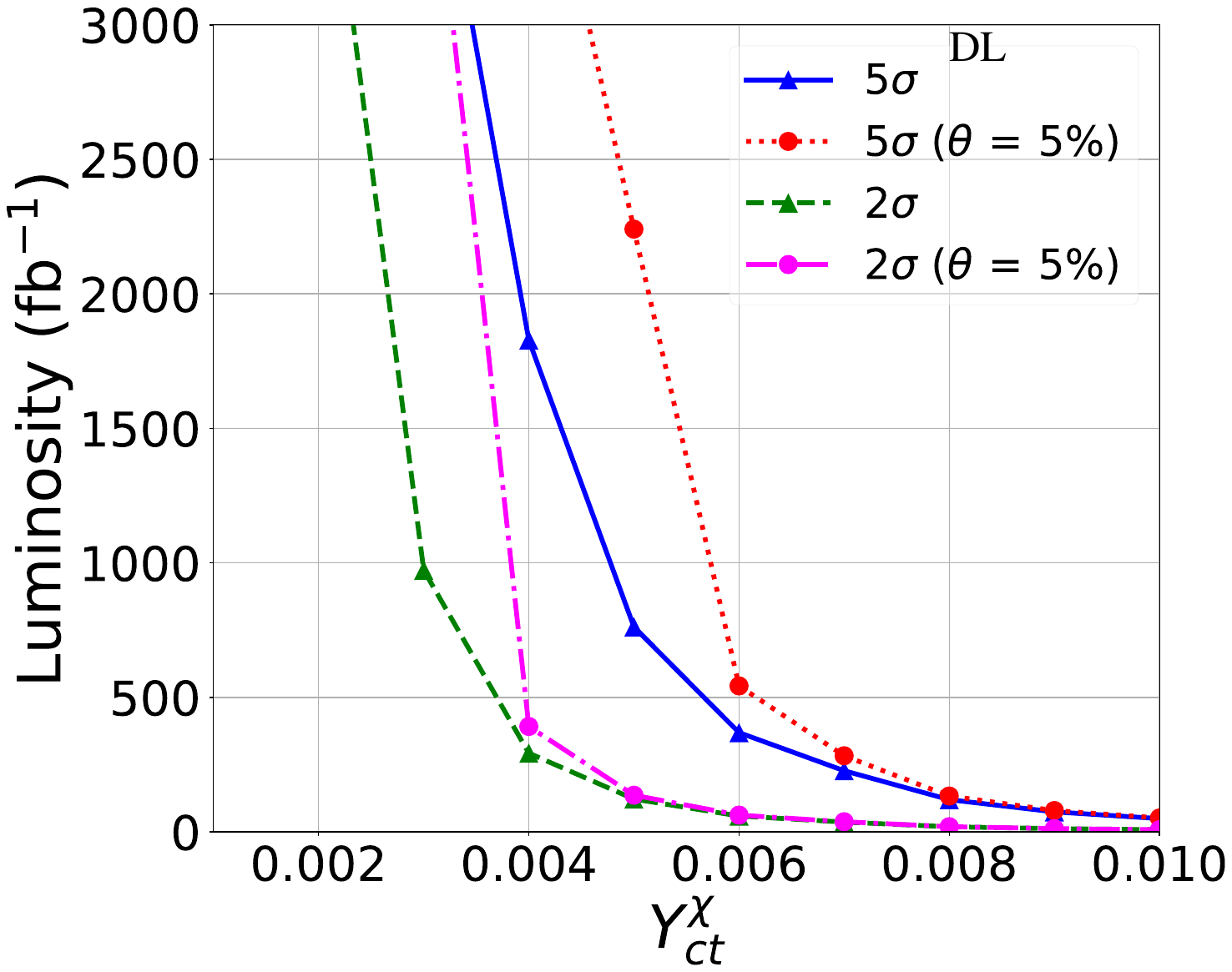}
  }
  \subfloat[SL : $m_\chi\,=\,20$ GeV]{
    \label{fig:SL20}
    \centering
    \includegraphics[width=0.45\textwidth]{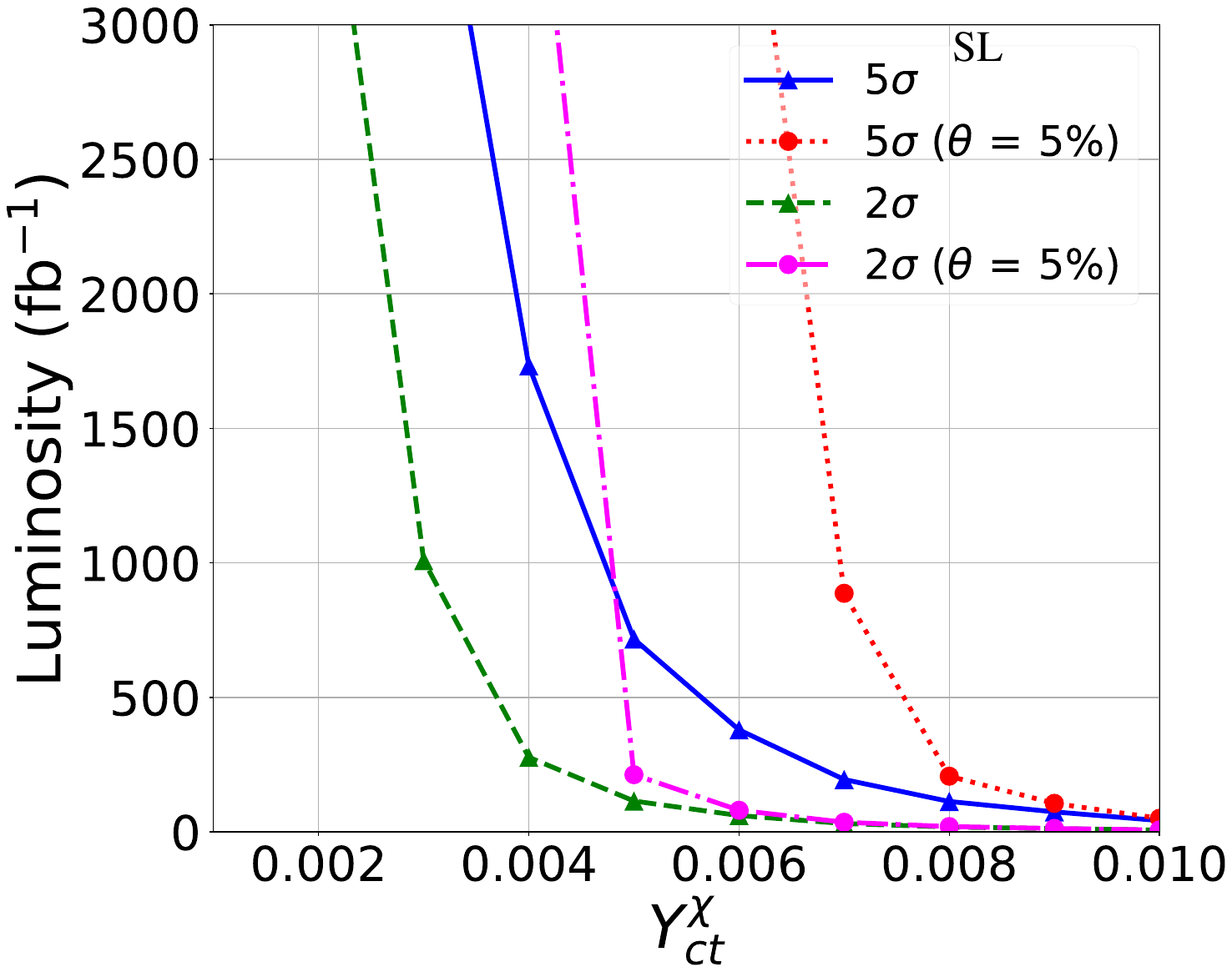}
  }
  \hspace{0.02\textwidth}
  \subfloat[DL : $m_\chi\,=\,60$ GeV]{
    \label{fig:DL60}
    \centering
    \includegraphics[width=0.45\textwidth]{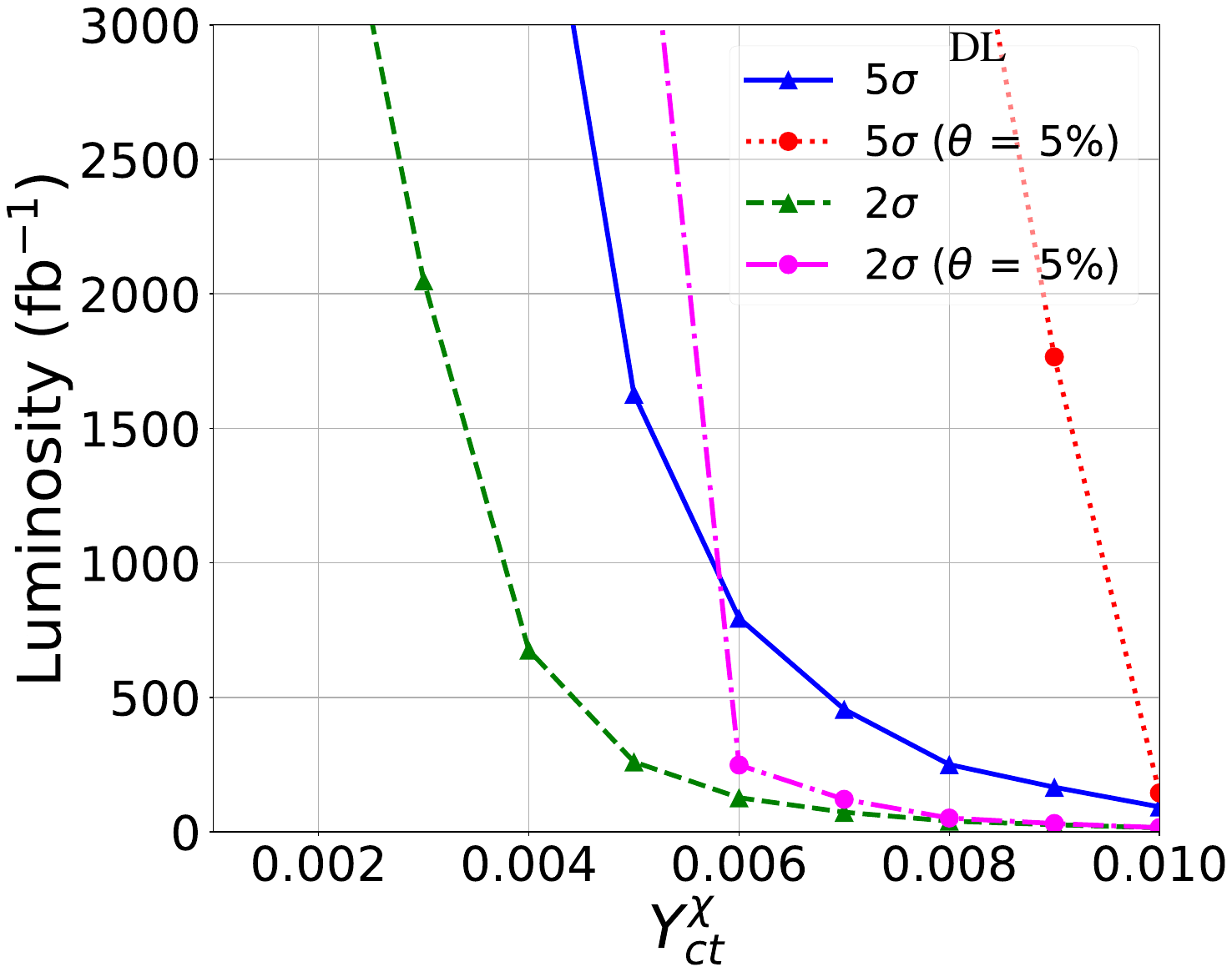}
  }
  \subfloat[SL : $m_\chi\,=\,60$ GeV]{
    \label{fig:SL60}
    \centering
    \includegraphics[width=0.45\textwidth]{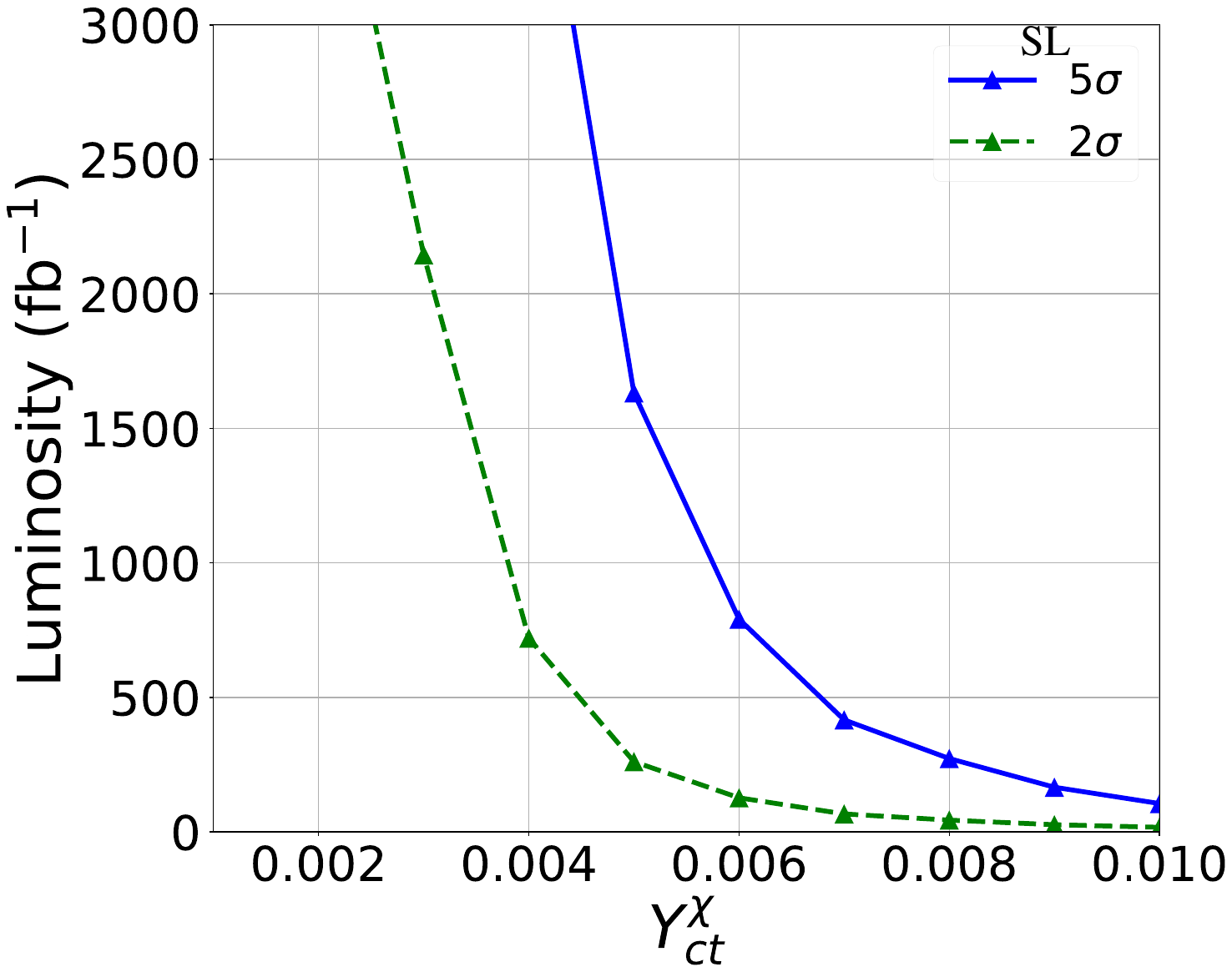}
  }
  \hspace{0.02\textwidth}
  \subfloat[DL : $m_\chi\,=\,100$ GeV]{
    \label{fig:DL100}
    \centering
    \includegraphics[width=0.45\textwidth]{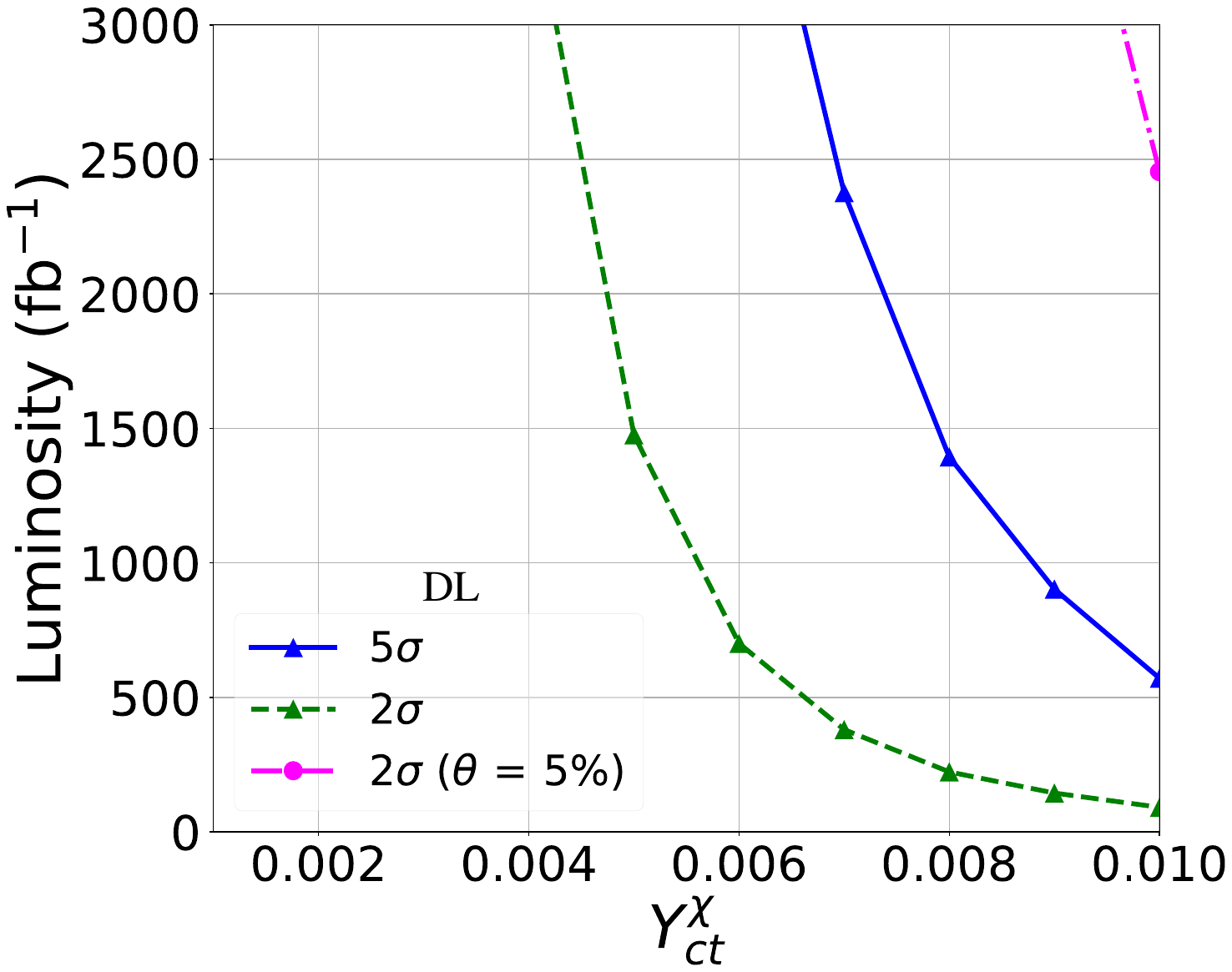}
  }
  \subfloat[SL : $m_\chi\,=\,100$ GeV]{
    \label{fig:SL100}
    \centering
    \includegraphics[width=0.45\textwidth]{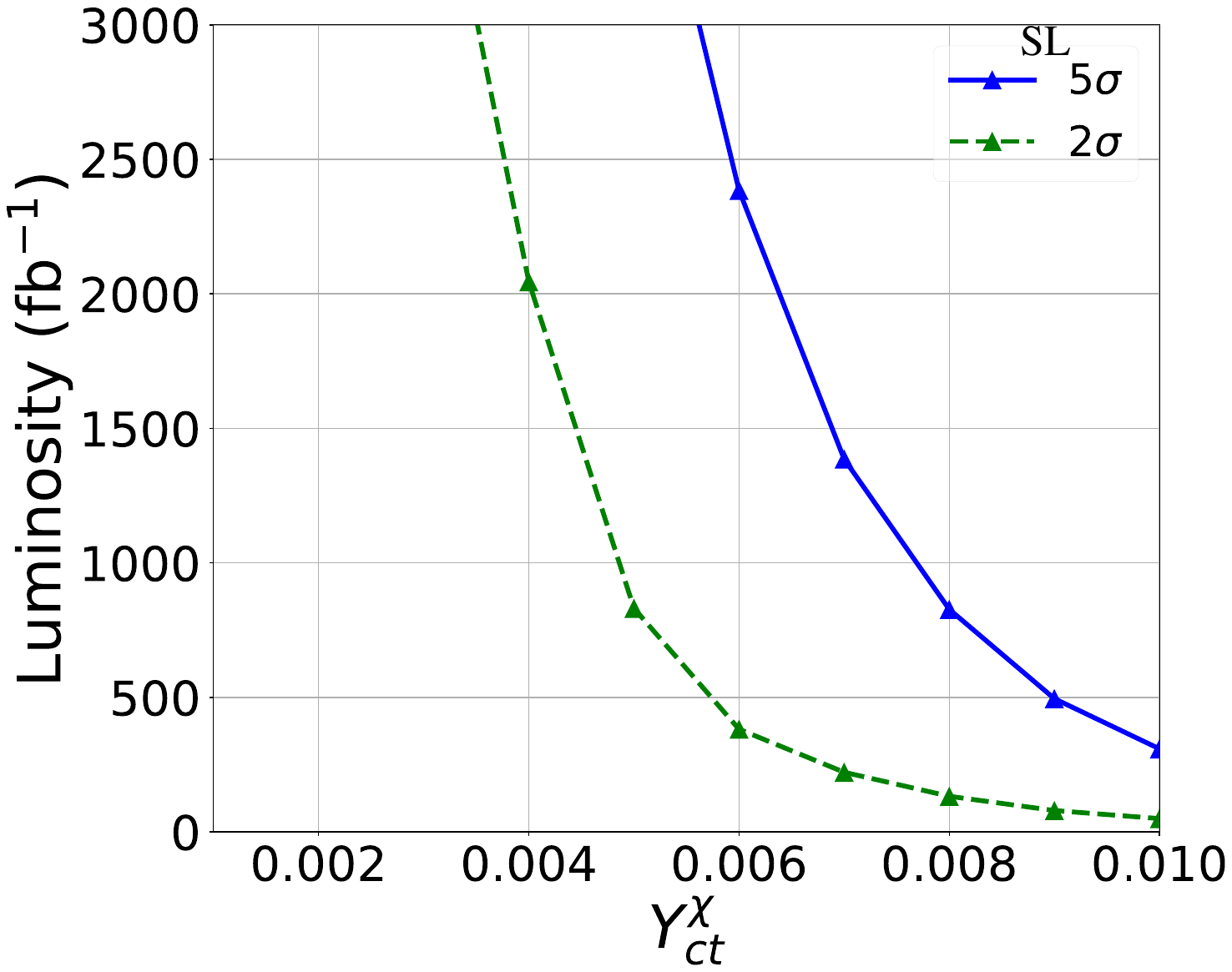}
  }
  \caption{\small \em Required integrated luminosity to achieve a
    $2\sigma$ exclusion and $5\sigma$ discovery. Figures of the left
    column correspond to DL channel and figures in the right column
    correspond to SL channel. Possible systematic uncertainties are
    expressed through $\theta$.}
  \label{fig:contours}
\end{figure}

Table \ref{tab:bestmva} basically describes the performances of three
analysis methods for $m_\chi = 60$ GeV in the DL channel. Here, we just
tabulate background rejections for four different signal efficiencies.
In addition, the right most column of Table \ref{tab:bestmva} shows the 
maximum significance at ${\cal L} = 3\,{\rm ab^{-1}}$.

Now we comment on the search prospects of the three benchmark points
at the HL-LHC. In Figure \ref{fig:contours}, we have drawn 2$\sigma$
and 5$\sigma$ contours in the integrated luminosity and $Y^\chi_{ct}$
plane with and without considering 5$\%$ linear-in-background
systematic uncertainty. In Figure \ref{fig:DL20} and \ref{fig:SL20},
for $m_\chi = 20$ GeV, the 2$\sigma$ and 5$\sigma$ contours drawn with
5$\%$ systematic uncertainty are shifted towards the higher values of
$Y^\chi_{ct}$, thus requiring more luminosity, for the SL channel
relative to the DL channel. Benchmarks with $m_\chi = 60$ GeV and
$m_\chi = 100$ GeV are accessible to $2 \sigma/5 \sigma$ significances
only with negligible systematic uncertainties. We thus conclude that
low $m_\chi$ provides promising avenues for exploration both for the
DL and SL channels at the HL-LHC. Increasing $m_\chi$ lowers the
search prospect.

\section{Summary and outlook} \label{Summary}

We have chalked out search strategies for a light exotic pseudoscalar
which has only off-diagonal Yukawa interaction with the quarks ($tj$)
and leptons ($\mu \tau$). For this, we have focused on the $t\bar{t}$
production and the subsequent decay channels at the 14 TeV HL-LHC with
different luminosities. Large accumulation of $t\bar{t}$ events at the
HL-LHC will be a gold mine to study such rare top quark decays. We
have adopted a simplified scenario with a minor reference to the big
picture based on flavor symmetry models that advocate such nonstandard
interaction of the exotic spin-0 states.

The present analysis is complementary to a previous study
\cite{Bhattacharyya:2022ciw} where the off-diagonal couplings of the
exotic scalar/pseudoscalar were assumed to involve only lighter
quarks. The present analysis focuses on off-diagonal Yukawa couplings
involving the top quark. It turns out that uncovering the signal
processes from the background is more tricky. We have employed three different Machine Learning techniques to perform multivariate analyses
and have found that DNN shows somewhat better performance compared to
the other two. No traditional method for hyperparameter 
tuning has been used.

We admit that our analysis does not take into consideration some
technical details, {\it e.g.}  jet faking as $\tau_h$ and/or
leptons, lepton charge misidentification, photon conversions into
lepton pairs, uncertainties on luminosity and trigger
efficiencies. Once the real data come, the ATLAS and CMS
experimentalists are urged to pursue deeper in this direction.

\section*{Acknowledgement}

We thank Siddharth Dwivedi, Ipsita Saha and Nivedita Ghosh for useful
discussions. IC acknowledges support from DST, India, under grant
number IFA18-PH214 (INSPIRE Faculty Award). TJ acknowledges the
support from Science and Engineering Research Board (SERB), Government
of India under the grant reference no. PDF/2020/001053. We acknowledge
support of the computing facilities of Indian Association for the
Cultivation of Science and Saha Institute of Nuclear Physics. Finally,
we thank the anonymous Referees for their insightful criticisms and
suggestions which have led to a substantial improvement of the
manuscript.

\bibliographystyle{JHEP}
\typeout{}
\bibliography{epjc.bib}

\end{document}